\documentclass[twocolumn]{aastex631}

\usepackage{natbib}

\usepackage{amsmath}

\usepackage[T1]{fontenc}

\graphicspath{{./}{figures/}}

\shorttitle{Proplyd Radio Recombination Lines}
\shortauthors{Boyden et al.}

\begin{document}


\title{{\bf {\large Discovery of Radio Recombination Lines from Proplyds in the Orion Nebula Cluster}}}

\correspondingauthor{Ryan Boyden}
\email{rboyden.astro@gmail.com}

\author[0000-0001-9857-1853]{Ryan D. Boyden}
\altaffiliation{Virginia Initiative on Cosmic Origins Fellow.}
\affiliation{Department of Astronomy, University of Virginia, Charlottesville, VA 22904, USA}
\affiliation{Space Science Institute, Boulder, CO 80301, USA}

\author[0000-0001-6527-6954]{Kimberly L. Emig}
\affiliation{National Radio Astronomy Observatory, Charlottesville, VA 22903, USA}

\author[0000-0002-4276-3730]{Nicholas P. Ballering}
\altaffiliation{Virginia Initiative on Cosmic Origins Fellow.}
\affiliation{Department of Astronomy, University of Virginia, Charlottesville, VA 22904, USA}
\affiliation{Space Science Institute, Boulder, CO 80301, USA}

\author[0000-0003-1413-1776]{Charles J. Law}
\altaffiliation{NASA Hubble Fellowship Program Sagan Fellow.}
\affiliation{Department of Astronomy, University of Virginia, Charlottesville, VA 22904, USA}

\author[0000-0002-9593-7618]{Thomas J. Haworth}
\affiliation{Astronomy Unit, School of Physics and Astronomy, Queen Mary University of London, London E1 4NS, UK}

\author[0000-0002-3389-9142]{Jonathan C. Tan}
\affiliation{Department of Astronomy, University of Virginia, Charlottesville, VA 22904, USA}
\affiliation{Department of Space, Earth and Environment, Chalmers University of Technology, SE-41296 Gothenburg, Sweden}

\author[0000-0003-2076-8001]{L. Ilsedore Cleeves}
\affiliation{Department of Astronomy, University of Virginia, Charlottesville, VA 22904, USA}
\affiliation{Department of Chemistry, University of Virginia, Charlottesville, VA 22904, USA}

\author[0000-0002-7402-6487]{Zhi-Yun Li}
\affiliation{Department of Astronomy, University of Virginia, Charlottesville, VA 22904, USA}



\begin{abstract}

{
We present new Atacama Large Millimeter/submillimeter Array observations that, for the first time, {detect} hydrogen and helium radio recombination lines from a protoplanetary disk. We {imaged} the Orion Nebula Cluster at 3.1 mm with a spectral setup that covered the $n=42 \rightarrow 41$ transitions of hydrogen (H41$\alpha$) and helium (He41$\alpha$). The unprecedented sensitivity of these observations {enables} us to search for radio recombination lines toward the positions of ${\sim}200$ protoplanetary disks. We detect H41$\alpha$ from 17 disks, all of which are HST-identified `proplyds.' The detected H41$\alpha$ emission is spatially coincident with the locations of proplyd ionization fronts, indicating that proplyd H41$\alpha$ emission is produced by gas that has been photoevaporated off the disk and ionized by UV radiation from massive stars. {We measure the fluxes and widths of the detected H41$\alpha$ lines and find line fluxes of ${\sim}30-800$ mJy km s$^{-1}$ and line widths of ${\sim}30-90$ km s$^{-1}$.} The derived line widths indicate that the broadening of proplyd H41$\alpha$ emission is dominated by {outflowing} gas motions {associated with external photoevaporation.} {The derived line fluxes, when compared with measurements of 3.1 mm free-free flux, imply that the ionization fronts of H41$\alpha$-detected proplyds have electron temperatures of ${\sim}6,000-11,000$ K and electron densities of ${\sim}10^6-10^7$ cm$^{-3}$.} Finally, we detect He41$\alpha$ towards one H41$\alpha$-detected source and find evidence that this system is helium-rich. Our study demonstrates that radio recombination lines are readily detectable in {ionized photoevaporating disks}, providing a new way {to measure disk properties} in clustered star-forming regions.
}

\end{abstract}


\section{{\bf Introduction}} \label{sec:intro}

Planets form in protoplanetary disks around young stars \citep[e.g.,][]{Keppler18}, 
and the properties of emerging planetary systems depend intimately on the structure, composition and evolution of protoplanetary disks \citep{Oberg21b, Oberg23}. The majority of stars, including the Sun, 
are born in {dense, massive} clusters 
\citep{Lada93, Lada03, Adams10, Krumholz19, Bergin23}. Understanding how disks evolve in clustered star-forming regions is crucial to interpreting the demographics of disks and exoplanets.

Theoretically, there is an expectation that the radiation environments of clusters influence disk properties \citep[for recent reviews, see][]{Parker20, Winter22}. Clusters host massive OB stars that irradiate their surroundings with UV photons. This intense radiation heats and ionizes protoplanetary disks in the cluster, driving material off their surfaces in the form of photoevaporative winds \citep[e.g.,][]{Johnstone98}.  
The external photoevaporation of disks is predicted to dominate over other internal and external mechanisms of disk dispersal 
in a range of cluster environments 
\citep[e.g.,][]{Scally01, Winter18, Winter20a, ConchaRamirez21, Coleman22, ConchaRamirez23, Coleman24, Gautam25}, 
leading to rapid disk truncation, {intense} 
mass loss, warmer disk temperatures, and shorter disk lifetimes in comparison with disks in lower-density star-forming regions \citep[e.g.,][]{Adams04, Clarke07, Tsamis13, Walsh13, Facchini16, Haworth18, Haworth19, ConchaRamirez19,  Haworth21a, Boyden23, Garate24, Ballering25}. 
If the extinction levels in clusters decline rapidly, such that disks become externally irradiated at early evolutionary stages 
\citep[e.g.,][]{Qiao22, Wilhelm23}, then external photoevaporation 
has the potential to influence both early- and late-stage planet formation \citep[e.g.,][]{Throop05, Haworth18b, Ndugu18, Winter22b}, or even suppress the formation of planets altogether \citep[e.g.,][]{Nicholson19, Parker21a, Qiao23, HuangS24}.

The Orion Nebula Cluster (ONC) provides one of the most compelling sites to study external photoevaporation.  
At a distance of the 400 pc \cite[][]{Hirota07, Kraus07, Menten07, Sandstrom07, Kounkel17, Grob18, Kounkel18}, it is the nearest and most readily observable high-mass cluster, comprising thousands of young (${1-2}$ Myr) low-mass (${<}2$ M$_{\odot}$) stars \citep{Hillenbrand97, Fang21} that are irradiated by the massive Trapezium stars, most notably the O6V star $\theta^1$ Ori C. 
In comparison with other nearby clusters, the ONC contains the largest population of disks with direct evidence {of ongoing} 
external photoevaporation. 
These disks, typically referred to as ``proplyds’’, consist of a central disk surrounded by a cocoon of ionized gas with a cometary morphology \citep[e.g.,][]{Odell94, Bally98, Ricci08}. 
The cometary morphologies of proplyds arise from material being photeovaporated off the disk surfaces by external UV radiation 
\citep{Johnstone98}. 
For most proplyds, the photoevaporative winds are driven by far-ultraviolet (FUV) radiation and then ionized by extreme-ultraviolet (EUV) radiation at an ionization front that is separated from the disk surface \citep[e.g.,][]{Storzer99, Richling2000}. 
For the proplyds closest to $\theta^1$ Ori C (i.e., within ${\lesssim}0.03$ pc), the EUV radiation is strong enough to overpower the FUV radiation and drive external photoevaporation, in which case the ionization front is closer to the disk surface than in the FUV-driven case \citep[e.g.,][]{Johnstone98}.

More than 200 proplyds have been identified in the ONC with optical \cite[e.g.,][]{Odell94, Bally98, Bally00, Ricci08, Aru24}, infrared \citep[e.g.,][]{Habart23, McCaughrean23}, 
{ millimeter \citep[e.g.,][]{Eisner08, Mann14}}, and radio imaging \citep[e.g.,][]{Churchwell87, Garay87, Zapata04a, Forbrich16, Sheehan16}. 
The detected proplyds span a range of morphologies, disk masses, and stellar properties,  though the majority appear to host compact disks with sizes smaller than a hundred au \citep[e.g.,][]{Eisner18, Boyden20, Otter21, Ballering23}. The prevalence, diversity, and proximity of the ONC’s proplyd population allows for a unique opportunity to constrain the relationships between external photoevaporation, star cluster evolution, and planet formation. 
Detailed spectroscopic studies at optical and infrared wavelengths have enabled constraints on the densities, temperatures, kinematics, and chemical compositions of photoevaporative winds for a handful of disk-proplyd systems \citep{Henney99, Henney02, Vasconcelos05, MesaDeldago12, Tsamis11a, Tsamis11b, MesaDeldago12, Tsamis13, MendezDelgado22, Kirwan23, Berne24, Goicoechea24, Zannese24}. 
However, to obtain a comprehensive picture of how external photoevaporation proceeds in a clustered star-forming environment,  
spectroscopic observations targeting  
larger and more diverse samples of photoevaporating disks are essential.

In this paper, we present new Atacama Large Millimeter/submillimeter Array (ALMA) observations that, for the first time, detect hydrogen and helium radio recombination lines from proplyds in the ONC. Radio recombination lines are electronic spectral line transitions associated with high principal quantum numbers;  
they are emitted when a free electron recombines with an ion and passes through high-quantum-number energy levels while cascading down to 
{lower energy levels} 
\citep{Gordon09, Draine11}. 
For decades, radio recombination lines have been used to measure the densities, temperatures, kinematics, and chemistry of 
{ {H\textsc{II}}}  regions \cite[see review in][]{Churchwell02}.  
With our deep, high-resolution ALMA observations, we can build upon studies of 
{H\textsc{II}} regions and use radio recombination lines to measure the physical conditions of a sample of 
photoevaporating protoplanetary disks.

\section{\bf {Observations and Data Reduction}} \label{sec:data}

\begin{figure*}[t!]
    \epsscale{1.0}
    \hspace{-1.0in}
    \plotone{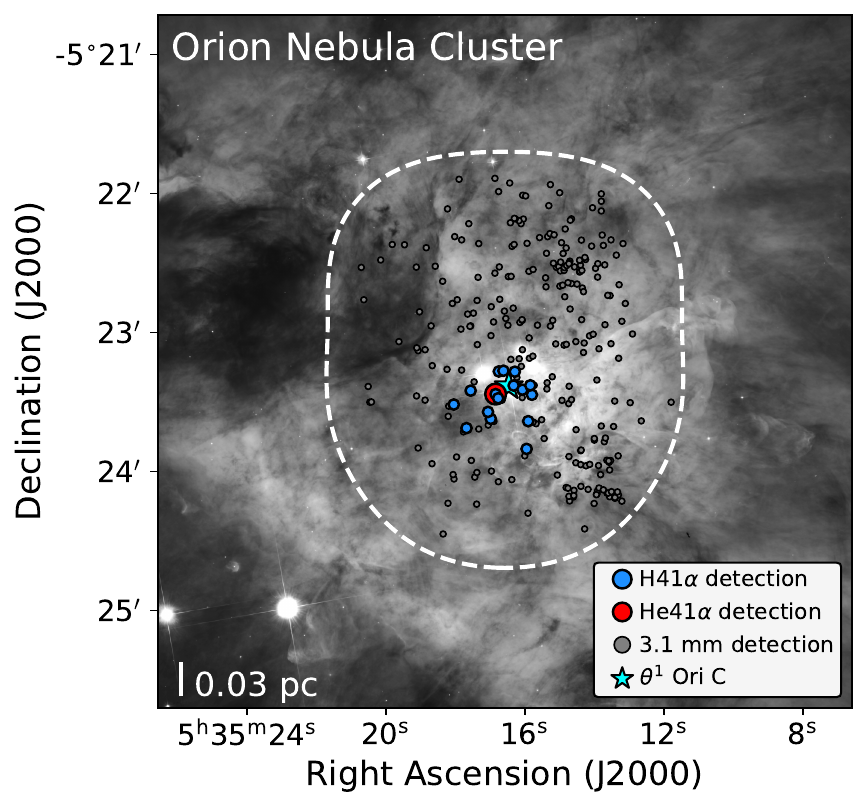}
    \vspace{-0.1in}
\caption{The Orion Nebula Cluster, as seen with HST/Advanced Camera for Surveys {\citep{Ricci08}}. The white dashed line depicts the field of view of our 3.1 mm ALMA mosaic. Blue circles indicate proplyds detected in H41$\alpha$. The red circle marks the proplyd that is also detected in  He41$\alpha$. Gray circles indicate the positions of 3.1 mm continuum detections that are not detected in H41$\alpha$ or He41$\alpha$. 
{ The cyan star indicates the position of the ionizing source $\theta^1$ Ori C.}  
\label{fig:ONC_spatial}
}
\end{figure*}

Our Cycle 6 ALMA Program mapped the ONC at Band 3 (3.1 mm). Observations were taken on 2019 July 7 and 2019 July 9 under project code 2018.1.01107.S. 
The C43-9 configuration was utilized to mosaic the {central  $2\rlap{.}'0 \times 2\rlap{.}'5$ of the ONC} with 10 pointings, covering baselines ranging from 149 m to 13.9 km
{with total on-source integration times of ${\sim}16$ minutes per pointing.} 
The spectral setup consisted of four windows centered at 90.5, 92.4, 102.5, and 104.4 GHz, with each window having a bandwidth of 1.875 GHz and channel spacing of 976.562 kHz { (${\sim} 3.5$ km s$^{-1}$)}. 
The Hydrogen $n = 42 - 41$ (H41$\alpha$) and Helium $n = 42 - 41$ (He41$\alpha$) recombination lines, with rest frequencies of 
{92.034 and 92.072 GHz,}
were covered in the 92.4 GHz window. 
\cite{Ballering23} presented the aggregate 3.1 mm continuum images produced from these ALMA observations, along with a full description of the continuum data reduction and imaging procedures. 
Below, we summarize our procedure for generating 
radio recombination line 
spectral cubes from these observations.

We began by retrieving the pipeline-calibrated measurement sets from the ALMA archive; data were downloaded and restored with the required pipeline version of CASA (5.4.0-70). We then used CASA version 6.5.4.9 for additional data processing. We used the {\tt uvcontsub} task to fit and remove the continuum with a first-order polynomial.
We then used the {\tt split} task to create standalone measurement sets for the continuum-subtracted 92.4 GHz spectral window, {as this window includes our spectral lines of interest (i.e., H41$\alpha$ and He41$\alpha$).} 

We imaged the continuum-subtracted measurement sets together
using the {\tt tclean} task with $specmode = cube$. 
We centered the image cube about the rest frequency of H41$\alpha$ and generated velocity channels in the range $\pm 200$ km s$^{-1}$. This velocity range was sufficient to cover emission from both H41$\alpha$ and He41$\alpha$, the latter of which, 
{at rest, is about $-122$ km s$^{-1}$ away}
from the H41$\alpha$ line.

All mosaic pointings were imaged together using the mosaic gridder with a phase center R.A. of 05:35:16.578 and decl. of --05:23:12.150, i.e., the same phase center used by \cite{Ballering23} for the continuum imaging.  To determine the positions of CLEAN boxes, we employed an iterative process where we first { placed $1'' \times 1''$ CLEAN boxes} around all objects detected above $8{\sigma}$, generated a CLEANed image, searched the residuals for detections above $4{\sigma}$, and then generated a new CLEANed image with additional { $1'' \times 1''$ CLEAN boxes} placed around additional detections. { Each CLEAN box was applied uniformly in all channels, and we employed a $2\sigma$ threshold to mitigate CLEANing of noise spikes.} { Imaging parameters were chosen to prioritize signal-to-noise over spatial resolution. For our final imaging}, we utilized a multiscale deconvolver with pixel scales of [0, 5, 15], natural weighting, and a {\it uv} taper of $0\rlap{.}''1$. { We did not perform the self-calibration and artifact removal procedure described in \cite{Ballering23}, which marginally improved the continuum rms in the vicinity of $\theta^1$ Ori A, BN, and Src I. These continuum sources do not produce strong artifacts in our continuum-subtracted line data cubes; we find that the cube rms is more spatially correlated with primary beam value than with proximity to bright continuum sources. }

{ Our final data products consist of two CLEANed data cubes, one binned into 3.5 km s$^{-1}$ channels (i.e., the approximate spectral resolution), and another binned into 10 km s$^{-1}$ channels. The coarser-binned data cube achieves higher S/N and is used to search for H41$\alpha$ and He41$\alpha$ detections (see Sections \ref{sec:results:H41a} and \ref{sec:results:He41a}).  
The finer-binned data cube is used to derive source properties from spectral line fitting (see Section \ref{sec:results:spectra}). 
In most regions of our cubes, the rms {noise in an individual velocity channel} is ${\sim}0.45 - 0.55$ mJy beam$^{-1}$ in our coarser-binned (10 km s$^{-1}$ channels) cube and ${\sim}0.65 - 0.75$ mJy beam$^{-1}$ in our finer-binned (3.5 km s$^{-1}$ channels) cube. In the outer regions of our cubes, the rms {increases} to values of ${\sim}1$ mJy beam$^{-1}$. 

The synthesized beam size for our cubes is about ${\sim}0\rlap{.}''2$. {This beam size is ${\sim}3$ times larger than the beam size achieved by \cite{Ballering23} for the 3.1 mm continuum image (${\sim}0\rlap{.}''07$).} At the {400 pc distance to Orion}, \cite[][]{Hirota07, Kraus07, Menten07, Sandstrom07, Kounkel17, Grob18, Kounkel18}, a beam size of ${\sim }0\rlap{.}''2$ corresponds to a spatial resolution of about $\sim$80 au. }

\begin{deluxetable*}{lllcchrrrc}
\tablenum{1}
\tabletypesize{\normalsize}
\tablewidth{0pt}
\tablecaption{Properties of Sources Detected in H41$\alpha$\tablenotemark{ }}\label{tab:source_properties}
\tablehead{ 
    \colhead{ID} & 
    \colhead{R.A.}  &
    \colhead{Decl.}  & 
    \colhead{$d_{\theta^1C}$}  & 
    \colhead{$D_{aperture}$}  & 
    \nocolhead{$F_{ff, 3.1mm}$} &  
    \colhead{Amp$_{H41\alpha}$} & 
    \colhead{$v_{cen,H41\alpha}$} & 
    \colhead{$\Delta v_{H41\alpha}$}  & 
    \colhead{$\int S_{H41\alpha} \ dV$}  \\ 
    \colhead{}        &  
    \colhead{(J2000)} & 
    \colhead{(J2000)}    & 
    \colhead{(pc)}    & 
    \colhead{($''$)}    & 
    \nocolhead{ (mJy)}   &    
    \colhead{(mJy)}  & 
    \colhead{(km s$^{-1}$)} & 
    \colhead{(km s$^{-1}$)} &
    \colhead{(mJy km s$^{-1}$)}  
}
\colnumbers
\startdata 
167-317 & 05:35:16.75 & --05:23:16.48 & 0.015 & 0.54 & $18.26 \pm 2.85$ & $11.36 \pm 0.66$ & $9.44 \pm 1.65$ & $58.15 \pm 3.88$ & $703 \pm 62$ \\
168-326 SE/NW\tablenotemark{a} & 05:35:16.84 & --05:23:26.29 & 0.013 & 0.6 & $19.29 \pm 2.01$ & $15.25 \pm 0.85$ & $15.97 \pm 1.24$ & $45.29 \pm 2.91$ & $735 \pm 63$ \\
163-317 & 05:35:16.29 & --05:23:16.58 & 0.013 & 0.48 & $9.09 \pm 1.40$ & $7.64 \pm 0.58$ & $10.94 \pm 1.67$ & $44.96 \pm 3.94$ & $366 \pm 42$ \\
158-323 & 05:35:15.84 & --05:23:22.47 & 0.018 & 0.51 & $9.45 \pm 1.52$ & $5.63 \pm 0.57$ & $16.33 \pm 2.57$ & $51.96 \pm 6.06$ & $311 \pm 48$ \\
161-324 & 05:35:16.07 & --05:23:24.38 & 0.012 & 0.42 & $4.99 \pm 0.80$ & $3.73 \pm 0.38$ & $3.92 \pm 2.86$ & $56.44 \pm 6.72$ & $224 \pm 35$ \\
168-328 & 05:35:16.77 & --05:23:28.08 & 0.014 & 0.33 & $3.17 \pm 0.56$ & $2.08 \pm 0.29$ & $16.60 \pm 2.44$ & $35.73 \pm 5.74$ & $79 \pm 17$ \\
163-323 & 05:35:16.32 & --05:23:22.57 & 0.004 & 0.36 & $3.06 \pm 0.71$ & $2.24 \pm 0.35$ & $-1.01 \pm 3.67$ & $48.65 \pm 8.64$ & $116 \pm 27$ \\
171-334 & 05:35:17.06 & --05:23:34.01 & 0.028 & 0.45 & $3.57 \pm 0.62$ & $2.77 \pm 0.41$ & $5.29 \pm 2.91$ & $40.10 \pm 6.85$ & $118 \pm 27$ \\
177-341W\tablenotemark{$\dagger$} & 05:35:17.68 & --05:23:40.97 & 0.050 & 0.9 & $11.51 \pm 0.22$ & $6.18 \pm 0.90$ & $2.37 \pm 3.18$ & $44.33 \pm 7.48$ & $292 \pm 65$ \\
158-327\tablenotemark{$\dagger$} & 05:35:15.79 & --05:23:26.56 & 0.021 & 0.75 & $9.85 \pm 0.13$ & $7.48 \pm 0.85$ & $12.10 \pm 2.20$ & $39.62 \pm 5.19$ & $316 \pm 55$ \\
159-338 & 05:35:15.90 & --05:23:37.96 & 0.033 & 0.66 & $2.14 \pm 0.32$ & $5.07 \pm 0.69$ & $12.44 \pm 2.39$ & $35.80 \pm 5.62$ & $193 \pm 40$ \\
180-331\tablenotemark{$\dagger$} & 05:35:18.05 & --05:23:30.81 & 0.049 & 0.63 & $5.91 \pm 0.11$ & $3.64 \pm 0.46$ & $22.67 \pm 5.36$ & $86.17 \pm 12.62$ & $334 \pm 65$ \\
170-337\tablenotemark{$\dagger$} & 05:35:16.98 & --05:23:37.05 & 0.032 & 0.45 & $7.04 \pm 0.24$ & $2.22 \pm 0.38$ & $-0.81 \pm 5.66$ & $67.58 \pm 13.34$ & $159 \pm 42$ \\
166-316 & 05:35:16.62 & --05:23:16.15 & 0.014 & 0.21 & $1.57 \pm 0.32$ & $0.76 \pm 0.15$ & $10.97 \pm 4.20$ & $42.99 \pm 9.89$ & $35 \pm 11$ \\
159-350\tablenotemark{$\dagger$} & 05:35:15.95 & --05:23:50.04 & 0.055 & 1.02 & $14.94 \pm 1.34$ & $8.73 \pm 1.17$ & $3.18 \pm 2.49$ & $38.01 \pm 5.87$ & $353 \pm 72$ \\
176-325 & 05:35:17.56 & --05:23:24.85 & 0.032 & 0.3 & $2.65 \pm 0.77$ & $1.02 \pm 0.20$ & $3.63 \pm 4.70$ & $49.70 \pm 11.06$ & $54 \pm 16$ \\
\enddata
\tablenotetext{ }{{ Notes.} 
Column (1): proplyd name. Columns (2) and (3): source coordinates. Column (4): projected distance from $\theta^1$ Ori C. 
Column (5): circular aperture diameters used to  
extract radio recombination line spectra. 
Columns (6), (7), and (8): best-fit H41$\alpha$ peak line flux, central velocity, and line width, derived by fitting Gaussian line profiles to the extracted spectra. Line widths are taken as the Gaussian full-width-at-half maximum. 
Column (9): Integrated H41$\alpha$ line flux, obtained by integrating over the best-fit Gaussian line profiles. 
}
\tablenotetext{a}{168-326 SE/NW  consists of two individual proplyds, 168-326 SE and 168-326 NW,  whose H41$\alpha$ emission is blended together in our $0\rlap{.}''2$ resolution data cubes. In Columns (2) and (3), we use the coordinates of 168-326 SE to indicate the position of 168-326 SE/NW.}
\tablenotetext{\dagger}{Indicates a proplyd whose disks and ionization fronts were spatially isolated in the 3.1 mm continuum by \cite{Ballering23}. }
\vspace{-0.4in}
\end{deluxetable*}

\begin{figure*}[ht!]
    \epsscale{1.2}
    \hspace{-0.3in}    \plotone{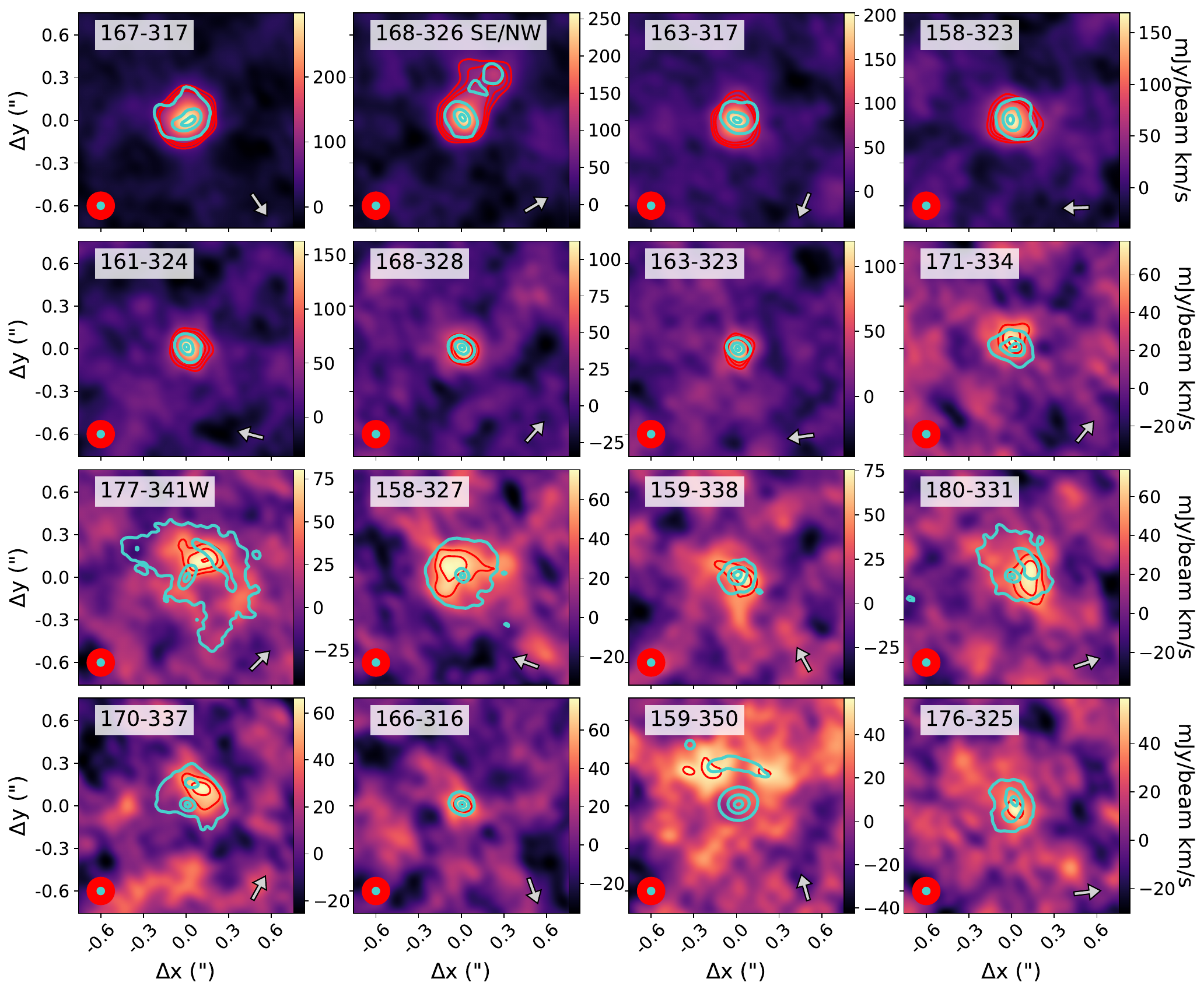}
     \vspace{-0.1in}
\caption{{Moment 0 (integrated intensity) maps of H41$\alpha$ for H41$\alpha$-detected proplyds in the ONC. Each panel corresponds to a  $1\rlap{.}"5 \times 1\rlap{.}"5$ ($600$ au $ \times$ $600$ au) region centered around each proplyd. 
Red contours 
show $3.5{\sigma}$, $4.5{\sigma}$, and $5.5{\sigma}$ 
H41$\alpha$ emission. 
Teal contours show 3.1 mm continuum emission from \cite{Ballering23}, with contours drawn at 10\%, 50\%,
and 90\% of the maximum continuum flux.
The synthesized beams for the 3.1 mm continuum (teal) and H41$\alpha$ (red) images are shown in the lower left corner of each panel. 
The name of each proplyd is specified in the top left corner of each panel. 
{ In the bottom right corner of each panel, we include an arrow that points to the direction of the ionizing source $\theta^1$ Ori C.}
{ All moment 0 maps are computed from our data cube with 10 km s$^{-1}$ channels. We generate moment 0 maps using 
velocity channels ranging from $0 - 60$ km s$^{-1}$ {with respect to the H41$\alpha$ rest frequency.
Standalone 3.1 mm continuum images for each  proplyd can be found in Appendix \ref{appendix:dust}.
}}
\label{fig:mom0_detections}}}
\end{figure*}

\

\ 

\ 

\ 

\  

\ 

\

\section{\bf {Results}}\label{sec:results}

\subsection{H41$\alpha$ detections}\label{sec:results:H41a}

To identify H41$\alpha$ emission from protoplanetary disks, we perform a detection search towards the positions of the 271 sources detected in the 3.1 mm continuum by \cite{Ballering23} and Ballering et al. in prep. { We use our 10 km s$^{-1}$ channel data cube to search for H41$\alpha$ detections}. {We consider H41$\alpha$ to be detected in the channel maps if emission within $0\rlap{.}''5$ of the known source position is detected {at} ${>}4{\sigma}$ in at least three consecutive velocity channels within $\pm 60$ km s$^{-1}$ of the H41$\alpha$ rest frequency. This velocity range more than covers the systemic velocities of ONC members. If H41$\alpha$ is not detected in the channel maps, then we compute moment 0 {(integrated intensity)} maps and spectra to see if it can be detected {at} ${>}4{\sigma}$ in a spatially or spectrally integrated frame. } {Our detection search is conducted manually using the search guidelines listed above. Our final detection count includes sources detected in the full channel maps as well as sources detected in the spatially and/or spectrally integrated frame only.}

\begin{figure*}[ht!]
    \epsscale{1.2}
    \hspace{-0.3in}    \plotone{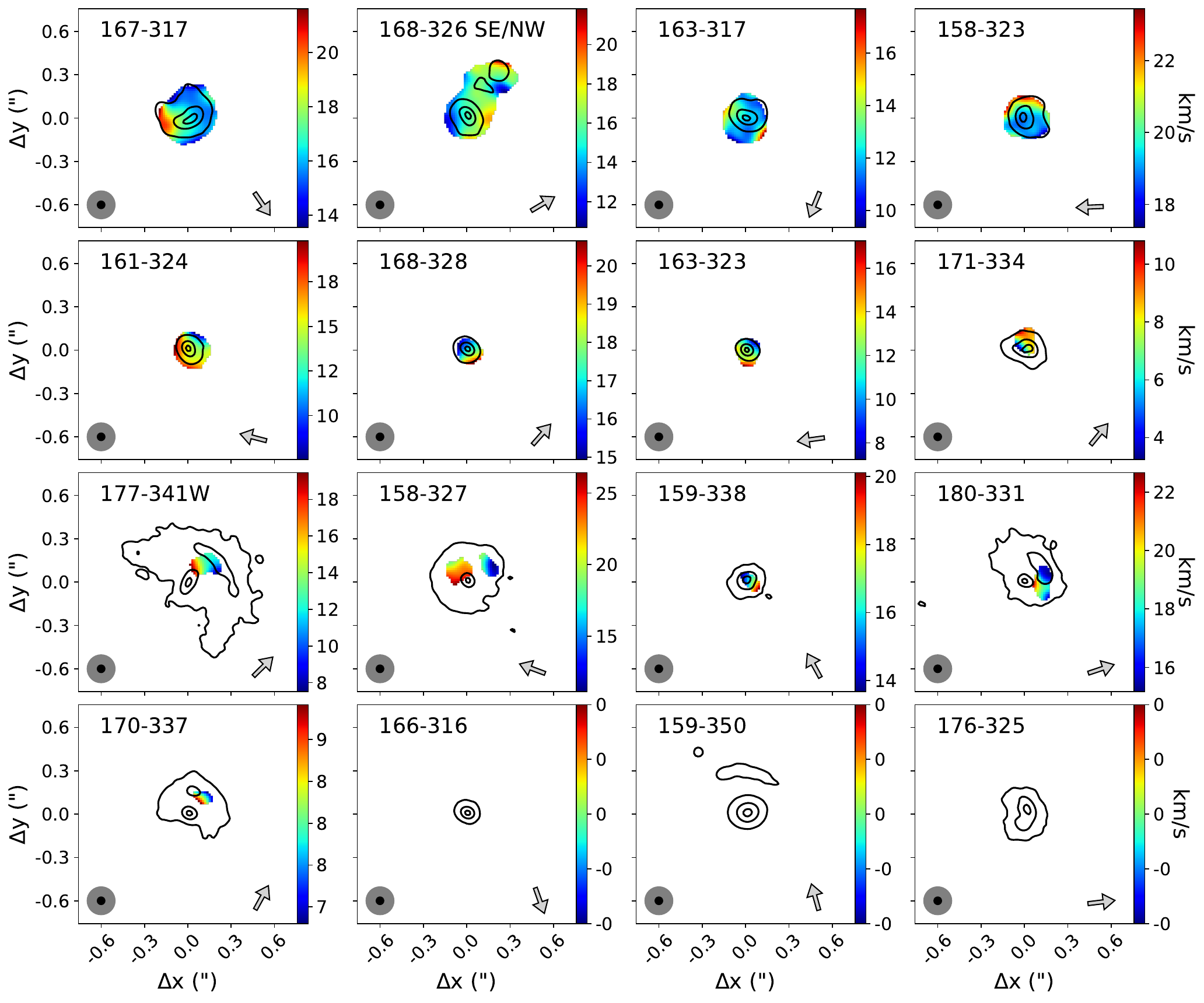}
     \vspace{-0.1in}
\caption{Moment 1 (intensity-weighted velocity) maps of H41$\alpha$ for the H41$\alpha$-detected proplyds, { generated from ${\geq}4.5{\sigma}$ channel map emission} over the same velocities used to create the moment 0 maps (see Figure \ref{fig:mom0_detections}). The layout of this plot to similar to that of Figure \ref{fig:mom0_detections}, except here we use black contours to show the 3.1 continuum emission, gray circles to show the synthesized beam for H41$\alpha$, and black circles to show the synthesized beam for the 3.1 mm continuum.
{ All moment 1 maps are generated from our data cube with 10 km s$^{-1}$ channels.}
\label{fig:mom1_detections}}
\end{figure*}

Table \ref{tab:source_properties} lists the coordinates and names of all sources detected in H41$\alpha$. 
We detect H41$\alpha$ emission from 17 protoplanetary disks. 
{ Of these 17 sources, 14 are detected above ${>}4{\sigma}$ in the channel maps, and 3 are detected above ${>}4{\sigma}$ in the moment 0 maps only (sources 166-316, 159-350, and 176-325). }
All 17 detections are HST-identified proplyds from the \cite{Ricci08} catalog. 
{ Five are proplyds that were presented in \cite{Ballering23} as 3.1 mm detections with spatially isolated disks and ionization fronts, and the rest are compact proplyds whose 
mm-continuum is dominated by free-free emission from the ionization front \citep[see example spectral energy distributions in][]{Mann14, Eisner18}.}
Two proplyds, 168-326 SE and 168-326 NW, are within $0\rlap{.}''3$ of each other and detected as a single object in our $0\rlap{.}''2$ resolution 
data cubes. 
For the remainder of this paper, we treat 168-326 SE and 168-326 NW as a single H41$\alpha$-detected source, which we refer to as 168-326 SE/NW. 
{ In Appendix \ref{appendix:dust}, we show  3.1 mm continuum sub-images of each H41$\alpha$-detected proplyd.}

\begin{figure*}[ht!]
    \epsscale{1.2}
    \hspace{-0.3in}    \plotone{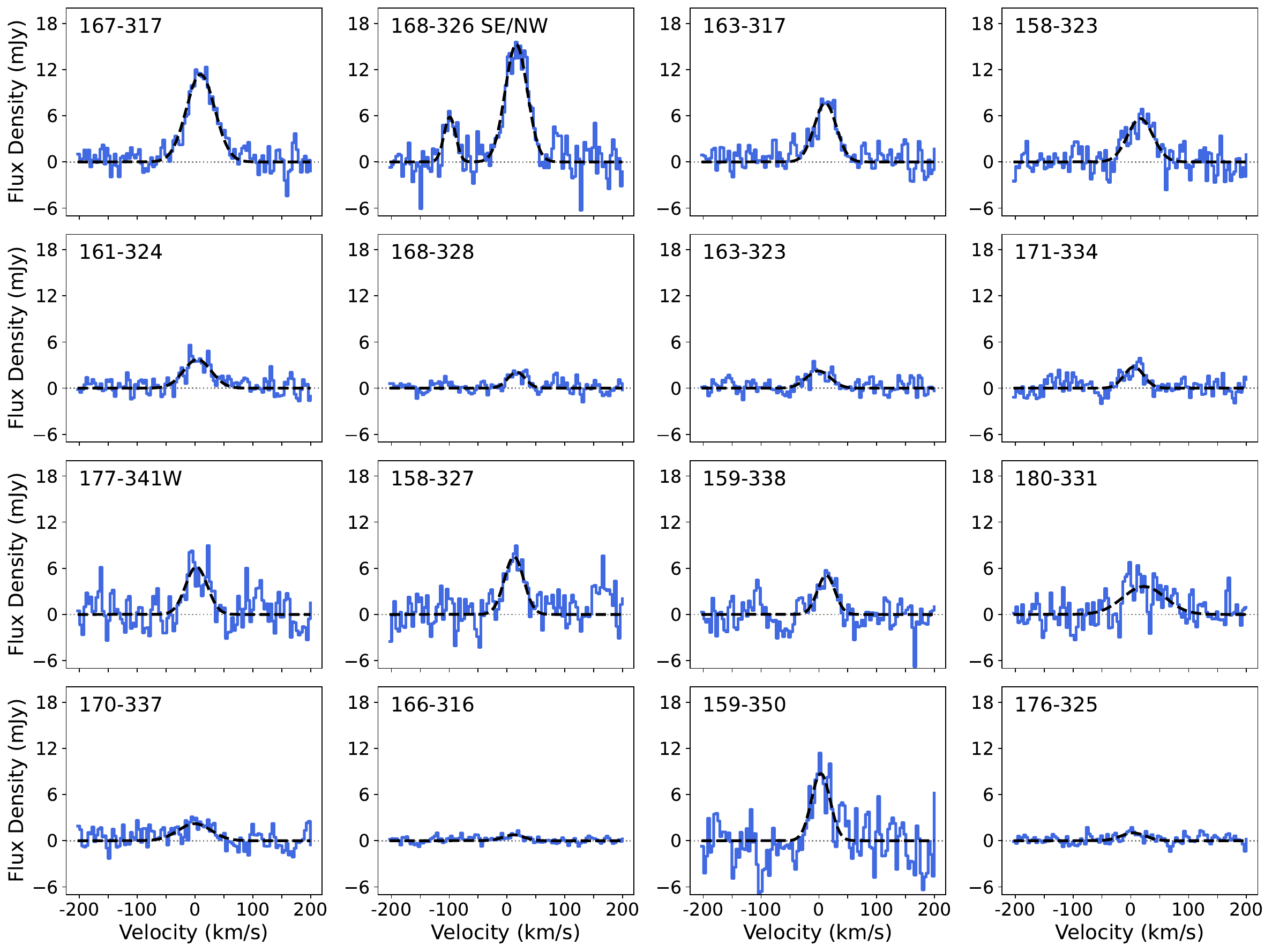}
     \vspace{-0.1in}
\caption{Radio recombination line spectra for H41$\alpha$-detected proplyds, { extracted from our data cube with 3.5 km s$^{-1}$ channels}. 
The spectra are centered about the rest frequency of the H41$\alpha$ line and constructed using circular apertures with diameters listed in Table \ref{tab:source_properties}. 
Solid blue lines show the computed spectra. 
Dashed black lines show best-fit Gaussian line profiles.
{Dotted gray lines show the zero point of the spectra.}
For 168-326 SE/NW, we fit two Gaussians in order to reproduce the detected He41$\alpha$ line at $v \approx -100$ km s$^{-1}$. 
{ Zoomed-in spectra for proplyds 168-328, 163-323, 166-316, and 176-325 are provided in Appendix \ref{appendix:Voigt}.} 
\label{fig:spec_detections}}
\end{figure*}

Figure \ref{fig:ONC_spatial} shows the locations of the H41$\alpha$-detected proplyds. The detections are concentrated in the center of the ONC near the massive Trapezium stars. Eleven are within ${\sim}0.03$ pc (in projected separation) of $\theta^1$ Ori C, where external photoevaporation is thought to be driven by EUV radiation rather than FUV radiation \citep{Johnstone98, Storzer99}. The other 5 detections have projected separations of ${\sim}0.03 - 0.06$ pc and are instead in regions where external photoevaporative is likely driven by FUV radiation. 
We note that the 5 detections with projected separations between $0.03$ and $0.06$ pc are all  in the same local region { to the south of 
$\theta^1$ Ori C.} 
These proplyds may be closer to $\theta^1$ Ori C and, thus, exposed to stronger UV fields than other proplyds with similar projected separations but different local positions in the ONC.

Figure \ref{fig:mom0_detections} shows moment 0 maps of H41$\alpha$ for the 
detected proplyds.  
The detected H41$\alpha$ emission is intimately associated with the locations of the 3.1 mm-identified ionization fronts.  
For the extended proplyds whose disks and ionization fronts are spatially separated in the 3.1 mm continuum (e.g., 177-341W, 158-327, 159-350), the H41$\alpha$ emission is concentrated towards the ionization front rather than the central disk. 
For the compact proplyds  
whose 
3.1 mm morphologies  
are dominated by the ionization fronts, 
(e.g., 167-317, 158-323, 161-234), the H41$\alpha$ emission is detected over the majority of positions where 3.1 mm continuum emission is seen.

{ Figure \ref{fig:mom1_detections} shows moment 1 {(intensity-weighted velocity)} maps of H41$\alpha$ for the detected proplyds. Many detections have H41$\alpha$ velocity gradients that are marginally detected at our current sensitivity and resolution. For proplyds 158-323, 161-324, 168-328, 163-323, 171-334, 177-341W, 159-338, and 170-337, the moment 1 maps show linear velocity gradients across the mm-identified ionization fronts. Photoevaporative winds can produce linear velocity gradients when the outflowing gas is rotating and/or reoriented along different flow directions while passing through an ionization front \citep[e.g.,][]{Richling2000, Haworth20}, so the detection of linear velocity gradients is consistent with the H41$\alpha$ emission being produced by an externally ionized disk wind. For proplyds 167-316, 168-326 SE/NW, 163-317, 158-327, and 180-331, the moment 1 maps reveal more complex features that deviate from a single velocity gradient. These deviations can be produced by photoevaporative disk winds with face-on orientations \citep[e.g.,][]{Haworth20}; however, they can also be explained by a combination of disk-wind and jet emission \citep[e.g.,][]{Klaassen18, Moscadelli21}. Finally, towards 166-316, 159-350, and 176-325, no clear velocity gradient is seen}.

Figure \ref{fig:spec_detections} shows spectra for the 
H41$\alpha$-detected proplyds. We extract spectra 
using circular apertures that are centered about the coordinates of each source. 
For 168-326 SE/NW, we place apertures around 168-326 SE and 168-326 NW { and sum the emission contained in both apertures}. 
Aperture sizes  
are determined on a source-by-basis and 
defined as the diameter that encapsulates all ${>}3{\sigma}$ H41$\alpha$ moment 0 emission associated with a source. We have verified that small (${\sim}0\rlap{.}''1$) adjustments to the aperture size have a negligible impact on the resulting spectra. 
In Table \ref{tab:source_properties}, we list the final aperture sizes used for each source. 
{ Typically, the brightest H41$\alpha$  detections have the most pronounced spectral amplitudes (e.g.167-317, 168-326 SE/NW); however, a couple of extended, low-surface-brightness detections have similar amplitudes as the bright H41$\alpha$ detections (e.g., 159-350).}

\subsection{He41$\alpha$ detections}\label{sec:results:He41a}

\begin{figure}[t!]
    \epsscale{1.2}
    \hspace{-0.15in}
    \plotone{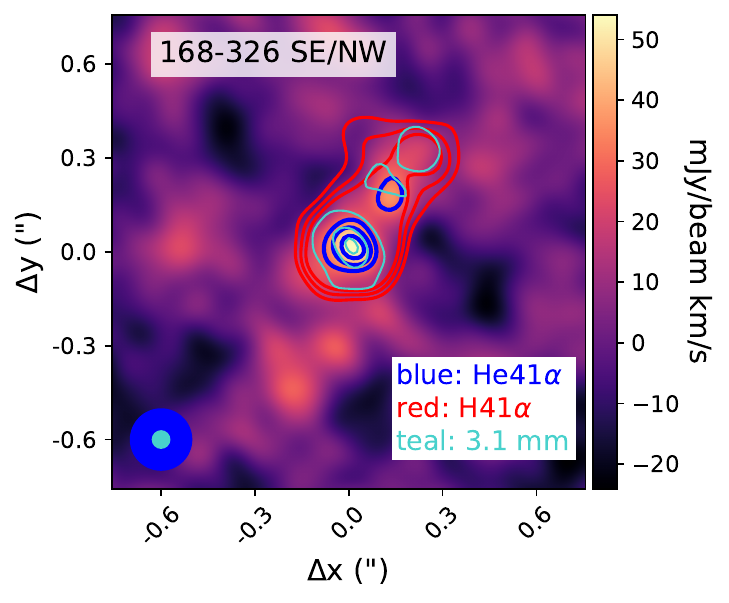}
     \vspace{-0.2in}
\caption{Moment 0 map of He41$\alpha$ for 168-326 SE/NW.  This moment map is generated from velocity channels in the range $-120$ to $-90$ km s$^{-1}$ {with respect to the H41$\alpha$ rest frequency}. Blue contours show $3.5{\sigma}$, $4.5{\sigma}$, and $5.5{\sigma}$ He41$\alpha$ emission. Teal contours show 3.1 mm continuum emission, with contours drawn at 10\%, 50\%,
and 90\% of the maximum continuum flux. Finally, red contours show $3.5{\sigma}$, $4.5{\sigma}$, and $5.5{\sigma}$ H41$\alpha$ emission (c.f., Figure \ref{fig:mom0_detections}). 
The synthesized beams for the 3.1 mm continuum (teal) and He41$\alpha$ (blue) images are shown in the lower left corner. The synthesized beam size is the same for the  He41$\alpha$ and H41$\alpha$  moment 0 maps. { All moment 0 maps are generated from our data cube with 10 km s$^{-1}$ channels.} \label{fig:mom0_He}
}
\end{figure}

{We detect He41$\alpha$ at ${>}4\sigma$ towards one H41$\alpha$ detected source: 168-326 SE/NW.} In Figure \ref{fig:spec_detections}, the detected He41$\alpha$ line is seen as a secondary peak at $v \approx 100$ km s$^{-1}$. Figure \ref{fig:mom0_He} shows a moment 0 map of He41$\alpha$ for 168-326 SE/NW. The detected He41$\alpha$ emission 
follows a similar morphology as the detected H41$\alpha$ emission. Both 168-326 NE and 168-326 SW appear to be detected in He41$\alpha$, but their emission is blended in our observations, similar to what is seen with H41$\alpha$. 

\subsection{Spectral line fitting}\label{sec:results:spectra}

We measure the {line fluxes, line widths, and central velocities} of the detected H41$\alpha$ and {He41$\alpha$} lines by fitting Gaussian line profiles to the observed spectra. 
For proplyd 168-326 SE/NW, 
{we fit two Gaussians}
in order to reproduce emission from both He41$\alpha$ and H41$\alpha$. 
For all other proplyds,   
{we fit a single Gaussian.}
{ All best-fit Gaussian parameters and their uncertainties are derived using the Levenberg-Marquardt procedure implemented in {\tt pyspeckit} \citep{Ginsburg22}.}
{Tables \ref{tab:source_properties} and \ref{tab:source_properties_He} list the best-fit peak line fluxes, line widths, central velocities, and integrated line fluxes for H41$\alpha$ and He41$\alpha$, respectively. 
Table \ref{tab:source_properties_He} also provides the integrated line flux ratio of He41$\alpha$ to H41$\alpha$ for 168-326 SE/NW.}

{ The detected H41$\alpha$ 
lines span an order of magnitude in flux, with peak line fluxes ranging from ${\sim}1$ to ${\sim}15$ mJy, and integrated line fluxes ranging from ${\sim}30$ to ${\sim}800$ mJy km s$^{-1}$.} The H41$\alpha$ central velocities tend to be around $0$ to ${\sim}20$ km s$^{-1}$, which is consistent with the local standard of rest velocities measured towards individual proplyds in the ONC \citep[e.g.,][]{Bally15, Factor17, Boyden20, Boyden23}. 
{For 168-328 SE/NW, the rest-frame H41$\alpha$ and He41$\alpha$ central velocities are within ${\sim}1\sigma$ of each other, and the integrated line flux ratio of He41$\alpha$ to H41$\alpha$ is ${\sim}0.2$, { 
which is about two times larger than the typical values seen in most, but not all, Galactic {H\textsc{II}} regions \citep[e.g.,][]{Churchwell74, Roshi17, GalvanMadrid24}.}} 
For the line widths, we typically derive full-width-at-half-maxima of ${\sim}40-60$ km s$^{-1}$ for H41$\alpha$ and ${\sim}20$ km s$^{-1}$ for He41$\alpha$. In general, these line widths are consistent with, but somewhat broader than, the line widths measured towards proplyds with optical recombination lines \citep[e.g.,][]{Henney99}. 
{ Finally, 180-331 has a best-fit H41$\alpha$ line width that is significantly broader than the typical values measured across our sample; though, at our current sensitivity and resolution, it is unclear whether this is driven by a low signal-to-noise ratio.}

{
While the detected H41$\alpha$ and He41$\alpha$ lines are fit well by Gaussians, we note that radio recombination lines can also exhibit Voigt line profiles when they are influenced by pressure broadening \citep[e.g.,][]{Keto08}. We have performed Voigt profile fitting to check whether any detected lines are fit better by Voigt than by Gaussian profiles. 
We find that the derived line profiles are qualitatively similar and  
yield similar reduced $\chi^2$ values (see example in Appendix \ref{appendix:Voigt}). 
However, at our current sensitivity and spectral resolution, we are typically unable to reliably constrain the Gaussian and Lorentzian components of the Voigt profiles, because in the velocity channels that are away from the line centers and where the line profiles differ most strongly, the signal-to-noise ratio is the lowest. 
We therefore opt to use Gaussian fitting to measure line properties, but note that higher sensitivity and spectral resolution are needed to firmly 
identify any 
non-Gaussian features (for a discussion on line broadening, see Section \ref{sec:width}).
}

\begin{deluxetable*}{hcccccc}
\tablenum{2}
\tabletypesize{\normalsize}
\tablewidth{0pt}
\tablecaption{
Measured He41$\alpha$ properties for 168-326 SE/NW
\tablenotemark{ }}\label{tab:source_properties_He}
\tablehead{ 
    \nocolhead{ID} &  
    \colhead{Amp$_{He41\alpha}$} & 
    \colhead{$V_{cen,He41\alpha^{\prime}}$} &  
    \colhead{$V_{cen,He41\alpha}$} &  
    \colhead{FWHM$_{He41\alpha}$}  & 
    \colhead{$\int S_{He41\alpha} \ dV$} & 
    \colhead{$y^+$} \\ 
    \nocolhead{}        &  
    \colhead{(mJy)} & 
    \colhead{(km s$^{-1}$)} & 
    \colhead{(km s$^{-1}$)} & 
    \colhead{(km s$^{-1}$)}  & 
    \colhead{(mJy km s$^{-1}$)}    & 
    \colhead{}    
}
\colnumbers
\startdata 
168-326 SE/NW & $5.82 \pm 1.17$ & $-98.94 \pm 2.34$ & $23.06 \pm 2.34$ & $23.71 \pm 5.52$ & $147 \pm 17$ & $0.200 \pm 0.023$ \\
\enddata
\tablenotetext{ }{{ Notes.} 
Column (1): best-fit He41$\alpha$ peak line flux.
Column (2): best-fit He41$\alpha$ central velocity with respect to the H41$\alpha$ rest frequency. 
Column (3): best-fit He41$\alpha$ central velocity with respect to the He41$\alpha$ rest frequency. 
Column (4): best-fit He41$\alpha$ line width, computed as the Gaussian full-width-at-half maximum.
Column (5): Integrated He41$\alpha$ line flux, derived by integrating over the best-fit Gaussian line profiles. 
Column (6): abundance of singly ionized helium relative to ionized hydrogen, computed as the ratio of the measured He41$\alpha$ and H41$\alpha$ line fluxes.
}
\end{deluxetable*}

\subsection{Electron temperatures}\label{sec:discussion:Te}

Radio recombination lines are powerful tools for measuring the temperature of ionized gas. 
At millimeter and centimeter wavelengths, the excitation conditions of hydrogen recombination lines are dominated by collisions with ionized plasma \citep{Gordon09, Draine11}. 
The emerging line intensities relate directly to the electron gas temperature via the Boltzmann equation, and the system is well approximated by local thermodynamic equilibrium (LTE).  

Here we use measurements of H$41\alpha$ emission and 3.1 mm free-free emission to calculate the electron temperature of each H$41\alpha$-detected proplyd. 
As discussed in Appendix A of \cite{Emig20}, the integrated line flux of an $n>40$ hydrogen recombination line with $\Delta n = 1$ can be expressed as: 
\begin{multline}\label{eq:int_SL}
    \int S_n dv =  (65.13 \textrm{ mJy km s}^{-1}) b_{n+1} \Big(\frac{n_e n_p V}{5 \times 10^{8} \textrm{ cm}^{-6} \textrm{ pc}^{3}}\Big) \\  \times \Big(\frac{D}{3.8 \textrm{ Mpc}}\Big)^{-2} \Big(\frac{T_e}{10^{4} \textrm{ K}}\Big)^{-1.5}
    \Big(\frac{\nu}{100 \textrm{ GHz}}\Big),
\end{multline} 
{where $\int S_n dv$ denotes the integrated line flux of a hydrogen radio recombination line transition to quantum number $n$,}
$b_{n+1}$ is the non-LTE departure coefficient (for LTE, $b_{n+1} = 1$), $n_e$ is the electron density, $n_p$ is the hydrogen gas density, $V$ is the source volume, $D$ is the source distance, $T_e$ is the electron temperature, and $\nu$ is the rest frequency of the hydrogen recombination line. 
 Moreover, in the optically thin regime \citep[as expected for proplyd free-free emission at 3.1 mm; e.g., ][]{Sheehan16, Boyden24}, 
 the {free-free flux density} can be written as 
\begin{multline}\label{eq:FF_thin}
    S_{ff} =  (2.080 \textrm{ mJy}) Z^{1.882} \Big(\frac{n_e n_+ V}{5 \times 10^{8} \textrm{ cm}^{-6} \textrm{ pc}^{3}}\Big) \\  \times \Big(\frac{D}{3.8 \textrm{ Mpc}}\Big)^{-2} \Big(\frac{T_e}{10^{4} \textrm{ K}}\Big)^{-0.323}
    \Big(\frac{\nu}{100 \textrm{ GHz}}\Big)^{-0.118},
\end{multline}
{where $S_{ff}$ is the free-free flux}, $n_+$ is the density of ions, and $Z = 1$ is the effective nuclear charge. 

\begin{deluxetable*}{lrrrhcccr}
\tablenum{3}
\tabletypesize{\normalsize}
\tablewidth{0pt}
\tablecaption{Derived Proplyd Properties Based on Measurements of 3.1mm Free-free emission and H41$\alpha$ Emission \tablenotemark{ }}\label{tab:source_properties_derived}
\tablehead{ 
    \colhead{ID} &  
    \colhead{$F_{cont,3.1mm}$} & 
    \colhead{$F_{dust, 3.1mm}$} &
    \colhead{$F_{ff, 3.1mm}$} &  
    \nocolhead{$\int S_{H41\alpha} \ dV$}  & 
    \colhead{$R_{LC}$} & 
    \colhead{$r_{IF}$} &
    \colhead{$T_{e}$} &
    \colhead{$n_{e}$}\\ 
    \colhead{}        &  
    \colhead{(mJy)} & 
    \colhead{(mJy)} & 
    \colhead{(mJy)}    & 
    \nocolhead{(mJy km s$^{-1}$)}    & 
    \colhead{(km s$^{-1}$)}    & 
    \colhead{(au)}    &
    \colhead{(K)} &
    \colhead{($\times 10^6$ cm$^{-3}$)} 
}
\colnumbers
\startdata 
167-317 & $23.39 \pm 0.04$ & $5.13 \pm 2.85$ & $18.26 \pm 2.85$ & $703 \pm 62$ & $39 \pm 7$ & $32.5 \pm 2.3$ & $7100 \pm 1100$ & $12.62 \pm 2.18$ \\
168-326 SE/NW & $21.69 \pm 0.03$ & $2.40 \pm 2.00$ & $19.29 \pm 2.01$ & $735 \pm 63$ & $38 \pm 5$ & $20.0 \pm 1.2$ & $6600 \pm 800$ & $16.33 \pm 2.24$ \\
163-317 & $12.02 \pm 0.04$ & $2.93 \pm 1.40$ & $9.09 \pm 1.40$ & $366 \pm 42$ & $40 \pm 8$ & $23.0 \pm 0.1$ & $6900 \pm 1100$ & $13.60 \pm 2.50$ \\
158-323 & $13.00 \pm 0.03$ & $3.55 \pm 1.52$ & $9.45 \pm 1.52$ & $311 \pm 48$ & $33 \pm 7$ & $27.3 \pm 0.2$ & $8200 \pm 1500$ & $11.87 \pm 2.53$ \\
161-324 & $7.17 \pm 0.03$ & $2.18 \pm 0.80$ & $4.99 \pm 0.80$ & $224 \pm 35$ & $45 \pm 10$ & $19.7 \pm 0.1$ & $6300 \pm 1200$ & $13.19 \pm 2.83$ \\
168-328 & $4.70 \pm 0.03$ & $1.53 \pm 0.55$ & $3.17 \pm 0.56$ & $79 \pm 17$ & $25 \pm 7$ & $19.1 \pm 0.1$ & $10300 \pm 2400$ & $12.75 \pm 3.35$ \\
163-323 & $6.09 \pm 0.04$ & $3.03 \pm 0.71$ & $3.06 \pm 0.71$ & $116 \pm 27$ & $38 \pm 13$ & $16.4 \pm 0.1$ & $7200 \pm 2000$ & $17.57 \pm 5.52$ \\
171-334 & $5.19 \pm 0.02$ & $1.61 \pm 0.62$ & $3.57 \pm 0.62$ & $118 \pm 27$ & $33 \pm 9$ & $25.0 \pm 0.3$ & $8100 \pm 2000$ & $7.05 \pm 1.92$ \\
177-341W & $13.40 \pm 0.02$ & $1.89 \pm 0.22$ & $11.51 \pm 0.22$ & $292 \pm 65$ & $25 \pm 6$ & $124.0 \pm 0.4$ & $10200 \pm 1900$ & $2.69 \pm 0.58$ \\
158-327 & $10.82 \pm 0.03$ & $0.97 \pm 0.12$ & $9.85 \pm 0.13$ & $316 \pm 55$ & $32 \pm 6$ & $88.0 \pm 0.4$ & $8400 \pm 1200$ & $2.45 \pm 0.42$ \\
159-338 & $3.22 \pm 0.02$ & $1.08 \pm 0.32$ & $2.14 \pm 0.32$ & $193 \pm 40$ & $90 \pm 23$ & $25.6 \pm 0.5$ & $3500 \pm 800$ & $4.34 \pm 1.06$ \\
180-331 & $6.71 \pm 0.02$ & $0.80 \pm 0.10$ & $5.91 \pm 0.11$ & $334 \pm 65$ & $56 \pm 11$ & $88.0 \pm 0.4$ & $5200 \pm 900$ & $2.34 \pm 0.44$ \\
170-337 & $9.04 \pm 0.02$ & $1.99 \pm 0.24$ & $7.04 \pm 0.24$ & $159 \pm 42$ & $23 \pm 6$ & $88.0 \pm 0.4$ & $11200 \pm 2500$ & $4.18 \pm 1.05$ \\
166-316 & $2.44 \pm 0.04$ & $0.87 \pm 0.31$ & $1.57 \pm 0.32$ & $35 \pm 11$ & $22 \pm 8$ & $17.3 \pm 0.2$ & $11400 \pm 3500$ & $10.26 \pm 3.57$ \\
159-350 & $26.86 \pm 0.03$ & $11.92 \pm 1.34$ & $14.94 \pm 1.34$ & $353 \pm 72$ & $24 \pm 5$ & $144.0 \pm 0.4$ & $10800 \pm 2100$ & $2.44 \pm 0.53$ \\
176-325 & $4.44 \pm 0.01$ & $1.79 \pm 0.77$ & $2.65 \pm 0.77$ & $54 \pm 16$ & $20 \pm 8$ & $29.4 \pm 0.4$ & $12300 \pm 4300$ & $6.36 \pm 2.50$ \\
\enddata
\tablenotetext{ }{{ Notes.} 
Column (1): proplyd name. 
Column (2): 3.1 mm continuum flux density, computed over the same source aperture listed in Table \ref{tab:source_properties}.
Column (3): 3.1 mm dust continuum flux density, taken from \cite{Ballering23} and Ballering et al., in prep.
Column (4): free-free flux density at 3.1 mm, computed as the difference between the 3.1 mm total continuum flux and 3.1 mm dust continuum flux.
Column (5): Ratio of the integrated H41$\alpha$ line flux and 3.1 mm free-free flux.
{Column (6): Ionization front radius. 
For proplyds 177-341W, 158-327, 180-331, 170-337, and 159-350, we list the ionization front radii derived previously by \cite{Ballering23}. 
For all other sources, 
{ the ionization front radius is measured by fitting elliptical Gaussians to the high-resolution 3.1 mm continuum sub-images of each source (see Appendix \ref{appendix:dust}), where the ionization front radius is taken as the best-fit half-width-at-half maximum minor axis. }
For 168-326 SE/NW, the provided ionization front radius is the average of the values obtained for 168-326 SE and 168-326 NW.}
Column (7). Electron gas temperature, derived from 
Equation \ref{eq:Te_RLC}.
{Column (8). Electron density, derived from 
Equation \ref{eq:ne_ff}.}
}
\vspace{-0.2in}
\end{deluxetable*}

Since $\int S_n dv$ and $S_{ff}$ have similar dependencies on the electron density, volume, and source distance, their ratio 
is independent of these parameters. 
The ratio of  $\int S_n dv$ over $S_{ff}$
can therefore be expressed as 
\begin{multline}\label{eq:RLC}
    \frac{\int S_n dv}{S_{ff}} =  (31.31 \textrm{ km s}^{-1}) b_{n+1} \Big(\frac{T_e}{10^{4} \textrm{ K}}\Big)^{-1.177} \\ 
    \times \Big(\frac{n_p}{n_+}\Big) \Big(\frac{\nu}{100 \textrm{ GHz}}\Big)^{1.118}.
\end{multline}
{If we assume that the emission region is composed primarily of ionized hydrogen and helium, then 
\begin{equation}\label{eq:y_He}
    \frac{n_p}{n_+} = \frac{n_{H^+}}{n_{H^+} + n_{He^+}} = (1 + n_{H^+}/n_{He^+})^{-1} = (1 + y^+)^{-1},
\end{equation}
where $y^+$ denotes the abundance of singly ionized helium relative to ionized hydrogen.} Hence, we can derive the electron gas temperature via  
\begin{multline}\label{eq:Te_RLC}
    T_e =  (10^4 \textrm{ K}) \Big[ b_{n+1} (1 + y^+)^{-1} \\  
    \times \Big(\frac{R_{LC}}{31.31 \textrm{ km s}^{-1}}\Big)^{-1} \Big(\frac{\nu}{100 \textrm{ GHz}}\Big)^{1.118} \Big]^{0.85} ,
\end{multline}
where $R_{LC} = \frac{\int S_n dv}{S_{ff}}$. 
{For the remainder of this section, we refer to $R_{LC}$ as the ``line-to-continuum ratio.'' }

{Table \ref{tab:source_properties_derived} lists the 
3.1 mm free-free fluxes used to compute the line-to-continuum ratios. 
To measure the free-free flux of each proplyd,} we first compute a total 3.1 mm continuum flux over the same circular apertures used to construct the proplyd radio recombination line spectra (see Table \ref{tab:source_properties}). These apertures cover dust emission from the central disks in addition to free-free emission from ionized gas.  
To remove contributions from dust emission, we use the 3.1 mm dust continuum fluxes derived by \cite{Ballering23} and Ballering et al. in prep, 
as these were computed by spatio-spectrally isolating the central dust disk with {combined 0.87 mm, 3.1 mm, 1.3 cm, 3.6 cm, and 6 cm ALMA and 
NSF's Karl G. Jansky Very Large Array (VLA)
observations.}
{ The apertures used to measure dust continuum fluxes are comparable to the ones we use to derive continuum fluxes and radio recombination line spectra; we verified that small changes to the continuum apertures have a negligible effect on the derived 3.1 mm dust continuum flux.   }
In Table \ref{tab:source_properties_derived}, we 
list the derived dust continuum fluxes along with our 3.1 mm continuum and free-free fluxes. 

\begin{figure*}[ht!]
    \epsscale{1.2}
    \hspace{-0.2in}    \plotone{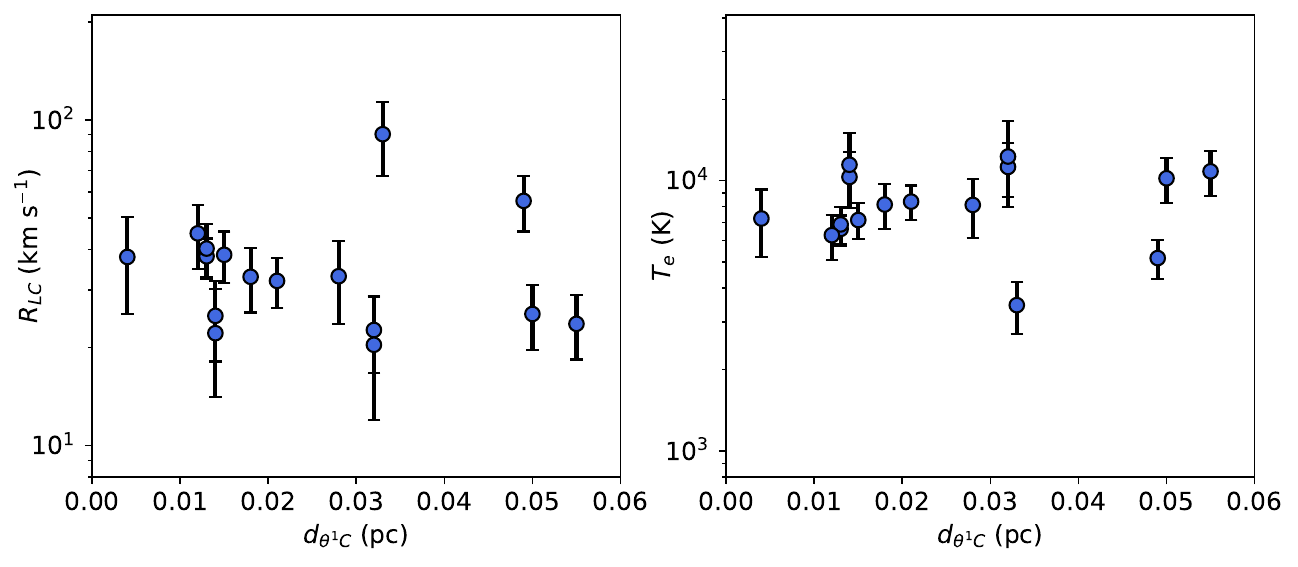}
     \vspace{-0.2in}
\caption{
Left: Ratio of integrated H41$\alpha$ line flux to 3.1 mm free-free continuum flux for H41$\alpha$-detected proplyds, plotted as a function of projection separation from $\theta^1$ Ori C. 
Right: proplyd electron gas temperature, derived from Equation \ref{eq:Te_RLC}, as a function of projected separation from $\theta^1$ Ori C. 
\label{fig:H41_RLC_Te}}
\end{figure*}

{
To derive proplyd electron temperatures from the line-to-continuum ratios, we assume LTE, in which case $b_{n+1} = 1$.
Our assumption of LTE is motivated by the departure coefficient calculations of \cite{Emig21}, 
{who find that for $n>40$},
$b_{n+1} > 0.8$ for the expected electron densities and temperatures of proplyds in the ONC \citep[$n_e >10^5$ cm$^{-3}$ and $T_e > 5000$ K; e.g.,][see also Section \ref{sec:discussion:ne}]{Churchwell87, Henney02, Ballering23}. 
If we were to assume a somewhat lower { departure} coefficient (e.g., $b_{n+1} = 0.8$), the derived electron temperatures would be ${\sim}1000-2000$ K lower than our computed values. 
}

{ 
For 168-326 SE/NW, we compute the electron temperature by taking $y^+$ as the ratio of the He41$\alpha$ and H41$\alpha$ line fluxes (see Table \ref{tab:source_properties_He}). Since hydrogen and helium radio recombination lines are optically thin \citep{Gordon09, Draine11}, their line ratios provide a direct measurement of  $y^+$. For all other H41$\alpha$-detected proplyds, we assume $y^+ = 0.1$, which is consistent with the typical values measured in Galactic {H\textsc{II}}  regions \cite[e.g.,][]{Churchwell74}, including the Orion Nebula \citep[e.g.,][]{Baldwin91, Pabst24}. If we were to assume a somewhat larger helium abundance that is similar to what we infer for 168-326 SE/NW (e.g., $y^+ = 0.2$), the derived electron temperatures would decrease by ${\sim}500-800$ K (for a discussion on helium abundance, see Section \ref{sec:discussion:He}).
}

Figure \ref{fig:H41_RLC_Te} plots the derived proplyd electron temperatures 
as a function of projected separation from $\theta^1$ Ori C. In Table \ref{tab:source_properties_derived}, we provide the derived electron temperatures and line-to-continuum ratios along with their uncertainties. 
The line-to-continuum ratios and electron temperatures are similar across our sample of H41$\alpha$-detected proplyds, indicating that 
{ionized gas from proplyds has a similar temperature}
over a range of disk, stellar, and environmental properties. 
Typically, we derive electron temperatures of ${\sim}7,000-9,000$ K. 
{These are broadly consistent with the expected electron temperatures of solar-metallicity {H\textsc{II}}  regions \cite[e.g.,][]{Haworth15, Wenger19, Balser24}. However, they suggest that many proplyds have { clumpy gas structures or have lower ionization front temperatures than the nominally assumed $10,000$ K value for externally--EUV-ionized photoevaporative winds \citep[e.g.,][]{Johnstone98}. }
}

\subsection{Electron densities}\label{sec:discussion:ne}

{Our ability to measure electron temperatures with radio recombination lines allows for 
a more accurate constraint of proplyd electron densities.} 
{For each H41$\alpha$-detected proplyd in our sample, we} 
compute a free-free optical depth from the peak 3.1 mm free-free flux density via $\tau_{ff} = I_{\nu} / B_{\nu} (T_e)$, where  $I_\nu$ denotes the peak intensity and $B_{\nu}$ denotes the Planck function. We then convert the free-free optical depth into an electron density under the assumption that the ionized gas is concentrated in a thin hemispherical shell at the ionization front. As shown in \cite{Ballering23}, this assumption yields the equation 
\begin{equation}\label{eq:ne_ff}
    n_e = \Bigg(\frac{EM^2 \sigma}{8 R_{IF}}\Bigg)^{1/3}
\end{equation}
where, $\sigma = 6.3 \times 10^{-18}$ {cm$^{2}$} is the ionization cross section; $R_{IF}$ is the radius at the proplyd ionization front; and EM is the emission measure, which relates directly to the free-free optical depth via Equation A.1b of \cite{Mezger67}: 
\begin{equation}
    \Big(\frac{EM}{\textrm{pc cm}^{-6}}\Big) = \Big(\frac{\tau_{ff}}{3.28 \times 10^{-7}}\Big) \Big(\frac{T_e}{10^{4} \textrm{ K}}\Big)^{1.35}
    \Big(\frac{\nu}{\textrm{GHz}}\Big)^{2.1}.
\end{equation} 
{We note that if the 
emitting gas layer is
thicker than we have assumed \citep[c.f.,][]{Henney98, Henney99},  
then the inferred electron densities would be ${\sim}5-10$ times lower than the values derived from a thin hemispherical shell geometry \citep[see discussion in][]{Boyden24}. Our current free-free-based electron densities may, thus, be upper limits that have been more robustly pinpointed with direct measurements of electron temperature. }

\begin{figure*}[ht!]
    \epsscale{1.2}
     \hspace{-0.15in}    \plotone{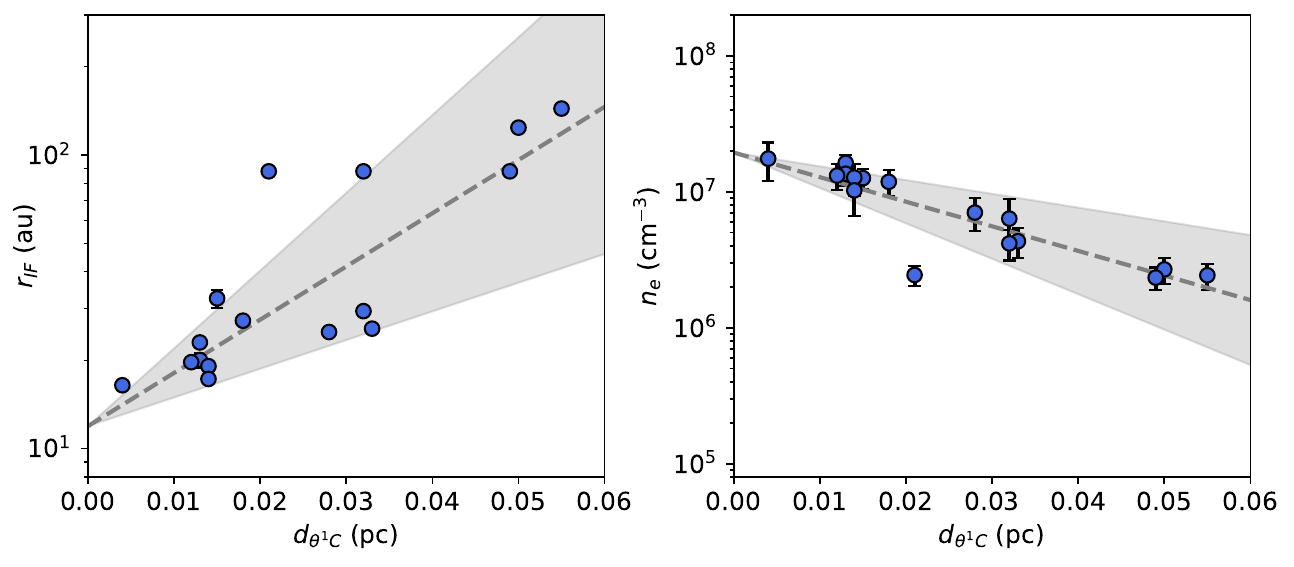}
 \vspace{-0.2in}
\caption{
Left: Proplyd ionization front radius as a function of projection separation from $\theta^1$ Ori C. { The fitted trend to the data is $\log_{10} (r_{IF} / AU) = (1.076 \pm 0.004) + (18.135 \pm 8.362) (d_{toc}/pc)$}. 
Right: proplyd electron density, derived from Equation \ref{eq:ne_ff}, as a function of projected separation from $\theta^1$ Ori C. The fitted trend to the data is $\log_{10}  (n_e / cm^{-3}) = (7.289 \pm 0.007) + (-18.081 \pm 7.962) (d_{toc}/pc)$. 
\label{fig:H41_ne}}
\end{figure*}

For the 5 H41$\alpha$-detected proplyds whose disks and ionization fronts are spatially isolated in 3.1 mm continuum, we use the 
ionization front radii from \cite{Ballering23} to compute the electron density. 
{These ionization front radii were computed by taking surface brightness cuts along the direction in which the disks and ionization fronts were detected and spatially isolated.}
{For the compact proplyds whose 3.1 mm morphologies are dominated by the ionization front, we measure their ionization front radii by fitting elliptical Gaussians to the high-resolution 3.1 mm  sub-images shown in Appendix \ref{appendix:dust}, 
where we take the best-fit half-width-at-half maximum minor axis as the ionization front radius. 
} 
Combining the measured ionization front radii with the measured 3.1 mm free-free fluxes and electron temperatures in Table \ref{tab:source_properties_derived}, 
we derive electron densities of ${\sim}10^6 - 10^7$ cm$^{-3}$.  In Table \ref{tab:source_properties_derived}, we list the computed ionization front radii and electron densities of each proplyd.

{ 
Figure \ref{fig:H41_ne} plots the  ionization front radii and electron densities of each H41$\alpha$-detected proplyd as a function of projected separation from $\theta^1$ Ori C. 
We identify a positive correlation between ionization front radius and projected separation \citep[consistent with previous studies; e.g.,][]{Ballering23, Aru24}. We also identify, for the first time, a negative correlation between electron density and projected separation. 
{Both correlations remain when we restrict our fitting to the compact proplyds whose ionization front radii are measured from Gaussian fitting (i.e., when we exclude proplyds whose ionization front radii were measured with surface brightness cuts).} 
Both trends can 
be explained by a reduction in EUV flux at increased projected separations, which 
enables photoevaporating gas to expand further away from the disk before becoming externally ionized \citep{Johnstone98}.
One proplyd, 158-327, has a lower electron density and a larger ionization front radius in comparison with other proplyds at similar projected separations. This suggests that 158-327 is further from $\theta^1$ Ori C than what is implied by its projected separation, or that trends with projected separation break down with increased sample sizes \citep[e.g.,][]{Parker21b}. 
}

\

\section{\bf {Discussion}}\label{sec:discussion}

\subsection{Line widths}\label{sec:width}

The H41$\alpha$ line widths that we derive for our sample of ONC proplyds are systematically larger than the line widths expected from pressure, thermal, {and turbulent} broadening. {For an {H\textsc{II}}  region, the line width of a radio recombination line can be written as   
\begin{equation}\label{eq:v_RRL}
    \Delta v_{RRL} = 0.543\Delta{v_p} +\Big[(\Delta v_t)^2 + (\Delta v_d)^2 + (0.4657 \Delta v_p)^2\Big]^{1/2} 
\end{equation}
\citep{Keto08}. Here, $\Delta v_{RRL}$ denotes the overall line width of a radio recombination line. $\Delta{v_p}$ denotes the Lorentzian width due to pressure broadening, which scales with the electron density, electron temperature, and principal quantum number as 
\begin{multline}\label{eq:v_pressure_actual}
    \Delta v_p = (1.43 \times 10^{-5} \textrm{ km s}^{-1})  \Big(\frac{n}{100}\Big)^{7.4} \Big(\frac{10^4 \textrm{ K}}{T_e}\Big)^{0.1} \\ 
    \Big(\frac{n_e}{10^4 \textrm{ cm}^{-3}}\Big)
    \Big(\frac{c}{\textrm{ km s}^{-1}}\Big),
\end{multline}
where $c$ denotes the speed of light \citep{Brocklehurst72}. $\Delta v_t$ denotes the Gaussian width due to thermal broadening, which, for a hydrogen recombination lines, scales with the electron temperature as 
\begin{equation}\label{eq:v_thermal}
    \Delta v_{t} = \sqrt{\frac{8 \ln 2 \ k_b T_e}{m_p}},
\end{equation}
where $k_b$ is the Boltzmann constant, and $m_p$ is the proton mass. Finally, $\Delta v_d$ denotes the Gaussian width due to dynamical broadening from turbulence and/or ordered gas motions.}

\begin{figure*}[ht!]
    \epsscale{1.2}
    \hspace{-0.2in}    
    \plotone{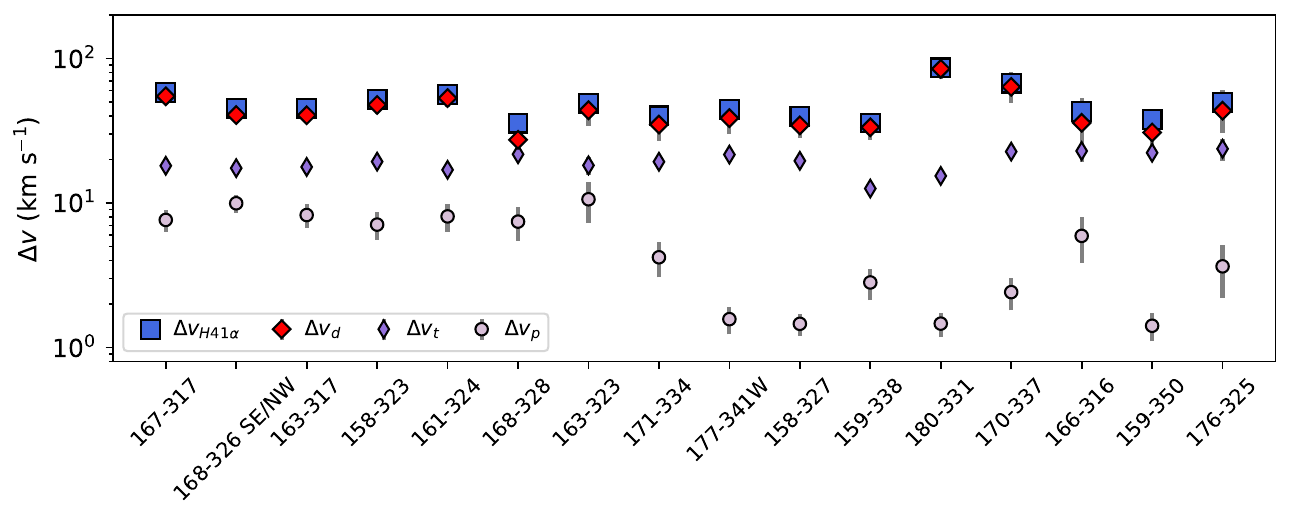}
     \vspace{-0.2in}
\caption{Measured H41$\alpha$ line width ($\Delta v_{H41\alpha}$) of each proplyd plotted along with the inferred line widths due to thermal broadening ($\Delta v_{t}$), pressure broadening ($\Delta v_{p}$), and dynamical broadening ($\Delta v_{d}$). 
\label{fig:H41_linewidths}}
\end{figure*}

{
Figure \ref{fig:H41_linewidths} plots the 
thermal and pressure line widths for each proplyd, computed via Equations \ref{eq:v_pressure_actual} and \ref{eq:v_thermal} using the electron densities and temperatures listed in Table \ref{tab:source_properties_derived}. 
Combining these with the measured H41$\alpha$ line widths in Table \ref{tab:source_properties}, we derive a dynamical line width, $\Delta v_d$, for each proplyd via Equation \ref{eq:v_RRL}, and and we show these in Figure \ref{fig:H41_linewidths}
along with the measured H41$\alpha$ line widths and inferred thermal and pressure line widths.

{ The line widths expected from pressure broadening are too low (${\sim}1 - 10$ km s$^{-1}$)  
to dominate the overall line broadening.}
Considering that our 3.1 mm continuum observations have detected the brightest and, thus, highest-electron-density ($n_e \approx 10^6 - 10^7$ cm$^{-3}$) proplyds in the ONC, it seems unlikely that any ONC proplyds are dense enough for pressure broadening to  
{ dominate}
the broadening of $n < 50$ recombination lines, 
{ though it may contribute for the highest-density proplyds. }
Moreover, while the thermal line widths (${\sim}20 - 25$ km s$^{-1}$) 
are broader than the line widths expected from pressure broadening, they are typically much lower than the measured H41$\alpha$ line widths. 

Figure \ref{fig:H41_linewidths} demonstrates that dynamical broadening provides the greatest contribution to the measured H41$\alpha$ line widths. 
In all cases, 
the inferred dynamical line widths are within $1{\sigma}$ of the measured H41$\alpha$ line widths. And, in the majority of cases, the dynamical widths exceed the thermal and pressure line widths by  ${>}2{\sigma}$. 
Since these dynamical widths are broader than the line widths expected from turbulence in {H\textsc{II}} regions 
\citep[${\sim}15$ km s$^{-1}$; e.g.,][]{Anderson11, Pabst24}, { we conclude that the dynamical and, thus, overall line broadening of proplyd H41$\alpha$ emission is dominated 
by { outflowing gas motions,} 
consistent with the presence of velocity gradients in Figure \ref{fig:mom1_detections}.
}}

The broad H41$\alpha$ line widths of proplyds 
can be explained  
{by an ionized photoevaporative wind.}
As gas photoevaporates off the disk 
and passes through an ionization front, pressure gradients cause the flow to accelerate from initial velocities of ${1-3}$ km s$^{-1}$ to supersonic values as high as ${\sim}30 - 60$ km s$^{-1}$ \citep[for {illustrative} examples, see][]{Richling2000}. {These expected terminal velocities match up well with the H41$\alpha$ line widths measured in our sample.}

Another possibility is that the broad H41$\alpha$ line widths are attributed to emission from both a disk wind and a jet. Stellar winds and jets associated with accretion tend to launch material at higher velocities than disk winds \citep[][and references therein]{Anglada18}. In high-mass protostars, jets produce the bulk of the hydrogen recombination line emission at high systemic velocities, though it is typically fainter than the disk emission seen at lower velocities. \citep[e.g.,][]{Prasad23, MartnezHenares24}. 
Two of the broadest-line-width proplyds in our sample\textemdash 167-317 and 170-337\textemdash are known jet hosts \citep{Bally98, Henney02, Tsamis11b, MendezDelgado21}. Moreover, 180-331
has a measured H41$\alpha$ line width that exceeds the typical 
${\sim}30-60$ km s$^{-1}$ velocities expected from external photeovaporation alone  (see Table \ref{tab:source_properties}). 
It seems plausible that the H41$\alpha$ line profile of 
180-331
is contaminated with high-velocity jet emission that, {at our current sensitivity and resolution}, cannot be distinguished from the photoevaporative wind. 
With {deeper, higher-resolution} radio recombination line observations, we can isolate disk wind versus jet emission while also taking census of jet-hosting proplyds in the ONC, as demonstrated in studies of high-mass protostars {\citep[e.g.,][]{Purser16, JiminezSerra20, Moscadelli21, GalvanMadrid23, Prasad23, MartnezHenares24}.}

Regardless of whether the H41$\alpha$ line widths are driven solely by disk winds or by disk winds and jets, our observations reveal that hydrogen radio recombination lines offer a new road-map for 
{ measuring}
the kinematics { and, by extension, mass-loss rates of ionized externally 
photoevaporating disks.} 
Higher angular resolution observations will enable 
{mapping}
of the flow { geometry and velocity} along different positions of the proplyd, allowing for more accurate determinations of the mass-loss rates, {rotational broadening}, and gas acceleration mechanisms. Millimeter- and submillimeter-band recombination lines appear especially suitable for mapping proplyd kinematics. At longer wavelengths, pressure broadening becomes significant, to the point where even for low-density photoevaporative flows, the line widths of recombination lines are dominated by pressure broadening (see Equation \ref{eq:v_pressure_actual}). 
{In future work, we intend 
to 
develop a framework for measuring the mass loss rates of proplyds with radio recombination lines.}

\subsection{Radio recombination lines as  thermometers for photoevaporative winds}

Our analysis demonstrates that hydrogen radio recombination lines provide a straightforward way to measure the 
{ temperatures of proplyd ionization fronts.} 
At optical wavelengths, which have so far been used to derive proplyd electron temperatures  \citep[e.g.,][]{Tsamis11a, MesaDeldago12, Tsamis13}, 
non-LTE effects dominate the gas excitation, and the electron densities and temperatures cannot be derived independently. 
Nebular background, saturation, and  reddening effects also introduce large systematic uncertainty to optical-based measurements of electron temperature, though in the case of proplyds, empirical correction techniques have been developed to mitigate these 
uncertainties \citep[e.g.,][]{Henney99, Tsamis11a, MesaDeldago12}.

With radio recombination lines, we can derive density-independent measurements of electron temperature that are unaffected by reddening, saturation, or the nebular background. 
{By detecting hydrogen and helium radio recombination lines from proplyds, we can also constrain ionized helium abundances to allow for a more robust determination of the electron temperature (see Equation \ref{eq:Te_RLC}).} 

{Radio-recombination-line based measurements of electron temperature are likely to achieve high precision when carried out at wavelengths longer than 3.1 mm,} particularly for compact proplyds where it is challenging to spatially separate the disk and ionization front. 
Currently, the precision on our derived electron temperatures is limited by the uncertainties on the 3.1 mm free-free fluxes, which are higher than the nominal continuum rms due to the spectral decomposition employed to separate out dust and free-free emission. Dust emission has a positive spectral index \citep{Hildebrand83}, {and at wavelengths between ${\sim}7$ and $20$ mm}, proplyd continuum flux measurements 
are expected to be dominated by optically thin free-free emission from the ionized wind \cite[e.g.,][]{Mann14, Sheehan16, Boyden24}.  {At these long wavelengths, hydrogen recombination lines remain in LTE down to densities as low as ${\sim}10^2$ cm$^{-3}$ and temperatures as low as ${\sim}1000$ K \citep[][]{Draine11, Emig21}. Long millimeter- and centimeter-based measurements of electron temperature should therefore be even more robust against systematic uncertainties associated with non-LTE effects.}

By obtaining accurate, high-precision electron temperatures with  radio recombination lines, we can more accurately measure the densities (e.g., see Section \ref{sec:discussion:ne}), chemical abundances, and mass-loss rates of ionized photoevaporative winds.  All of these quantities depend on the assumed electron temperature, with the chemical abundances derived from optical forbidden lines 
being especially sensitive to the electron temperature \citep[e.g.,][]{MesaDeldago12, MendezDelgado23}.

\subsection{Properties of H41$\alpha$ nondetections}

\begin{figure*}[ht!]
    \epsscale{1.2}
    \hspace{-0.3in}    \plotone{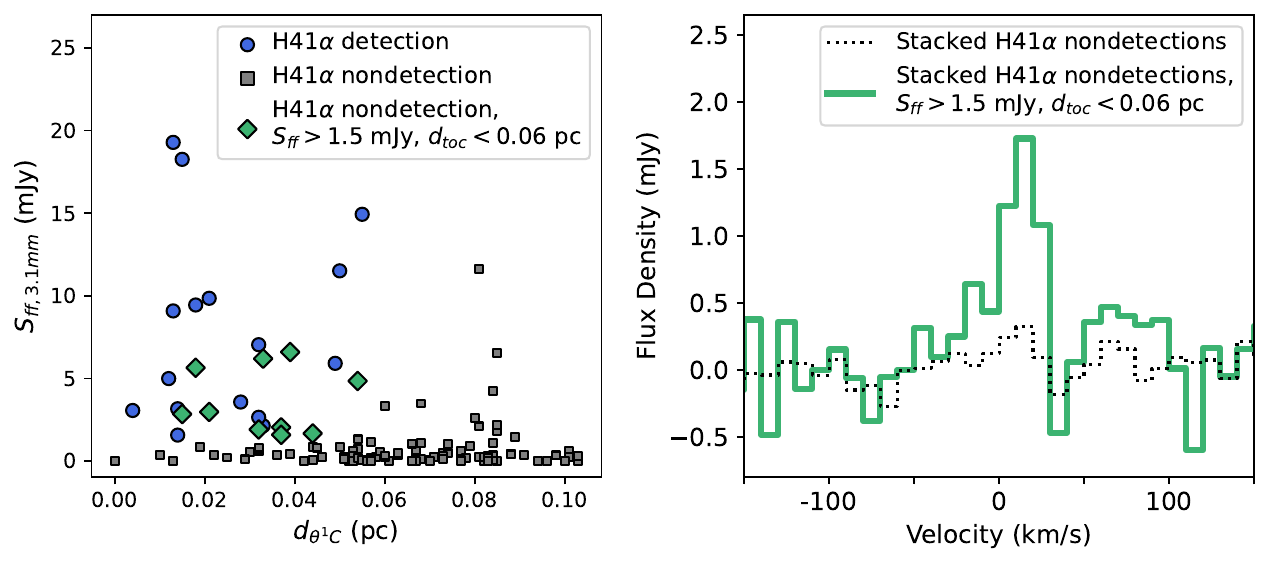}
     \vspace{-0.1in}
\caption{
Left: 3.1 mm free-free fluxes of ONC disks 
\citep{Ballering23}, plotted as a function of projection separation from $\theta^1$ Ori C. 
Blue circles indicate H41$\alpha$-detected proplyds. 
Green diamonds indicate H41$\alpha$ nondetections with 3.1 mm free-free fluxes ${>}1.5$ mJy and projected separations ${<}0.06$ pc (i.e., with similar free-free fluxes and projected separations as the H41$\alpha$ detections; see Tables \ref{tab:source_properties} and \ref{tab:source_properties_derived}). 
Grey squares indicate all other H41$\alpha$ nondetections. 
Right: Stacked average spectra for all H41$\alpha$ nondetections (dotted black line) and for the subset of H41$\alpha$ nondetections with 3.1 mm free-free fluxes ${>}1.5$ mJy and projected separations ${<}0.06$ pc (solid green line). { The stacked spectra are constructed using our data cube with 10 km s$^{-1}$ channels.}
\label{fig:H41a_nondetections}}
\end{figure*}

We perform a stacking analysis to explore the properties of ONC disks not detected in {H41$\alpha$.
The} left panel of Figure \ref{fig:H41a_nondetections} shows the 3.1 mm free-free fluxes of all mm-detected ONC disks in our maps as a function of projected separation from $\theta^1$ Ori C. The right panel shows stacked average radio recombination line spectra for the H41$\alpha$ nondetections. 
We compute the free-free fluxes and spectra of H41$\alpha$ nondetections by following the same procedures used for the H41$\alpha$ detections, except here we adopt a uniform circular aperture diameter of $0\rlap{.}''5$. 
Since the H41$\alpha$-detected proplyds tend to be free-free luminous and close to $\theta^1$ Ori C (see Figure \ref{fig:H41a_nondetections}),
we generate two stacked spectra, one for all ${\sim}$200 H41$\alpha$ nondetections, and another for the subset of nondetections with similar free-free fluxes ($S_{ff} > 1.5$ mJy) and projected separations ($d_{toc} < 0.06$ pc) as the H41$\alpha$ detections (see Figure \ref{fig:H41a_nondetections}).

When we restrict our stacking to the 10 nondetections with similar free-free fluxes and projections separations as the H41$\alpha$ detections, we see 
{a ${\sim}4 \sigma$ detection} of H41$\alpha$ in the stacked spectrum. 
The H41$\alpha$ emission of these sources appears to have just fallen below our detection threshold. 
Since free-free-bright H41$\alpha$ nondetections likely have similar electron densities as the H41$\alpha$ detections, we expect an H41$\alpha$ nondetection to, in this case, be attributed to either smaller line-to-continuum ratios or increased noise levels. 
We have examined the noise fluctuations in our ALMA data cube to check whether the rms {noise} is higher towards the positions of  free-free-bright H41$\alpha$ nondetections versus H41$\alpha$ detections. 
We find no significant differences in the local rms, 
indicating that 
free-free-bright H41$\alpha$ nondetections 
are driven by lower line-to-continuum ratios and, by extension,  warmer electron temperatures and/or lower gas metallicites 
(see Equation \ref{eq:RLC}).

When we stack all 
${\sim}200$ continuum sources in our maps, we see no H41$\alpha$ detection in the stacked spectrum, stacked moment maps, or stacked channel maps. 
{ This suggests that the average ONC disk has an H41$\alpha$ line flux that is ${\geq}10$ times fainter than our current {rms noise levels (i.e., ${<}0.5$ mJy)}. } 
For most disks, we expect this to be due to a lack of an ionization front,  
since only ${\sim}50\%$ of mm-detected ONC disks are HST-identified proplyds \citep[e.g.,][]{Eisner18}. 
But in cases where an H41$\alpha$ nondetection is a known proplyd with a fainter free-free flux than an H41$\alpha$-detected proplyd, 
we expect a factor of ${\sim}10$ lower H41$\alpha$ line flux to be attributed to a combination of warmer gas temperatures, lower gas metallicites, and/or lower electron densities (see Equations \ref{eq:int_SL} and \ref{eq:RLC}).

Equation \ref{eq:int_SL} suggests that a factor of ${\sim}3$ reduction in density reduces the H41$\alpha$ line intensity by a factor of ${\sim}10$. 
The majority of ONC proplyds may therefore have electron densities that are ${\sim}3$ times lower than the minimum electron density in our H41$\alpha$-detected proplyd sample, which would imply electron densities of ${<}10^6$ cm$^{-3}$ (c.f., Table \ref{tab:source_properties_derived}). {Moreover, if the electron density correlates inversely with distance from $\theta^1$ Ori C, as suggested from Figure \ref{fig:H41_ne}, this would imply that the majority of proplyds 
not detected in H41$\alpha$ are further from $\theta^1$ Ori C than what is implied by their projected separations.}

\begin{figure*}[ht!]
    \epsscale{1.1}
    \hspace{-0.3in}    \plotone{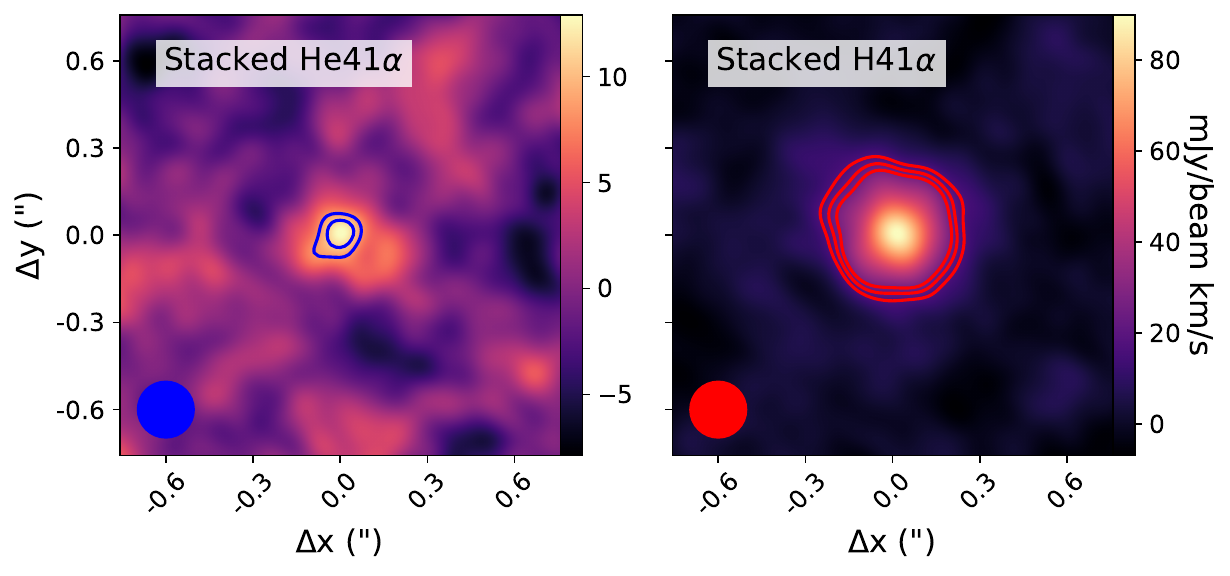}
     \vspace{-0.07in}
\caption{Stacked He41$\alpha$ moment 0 map (left panel) and H41$\alpha$ moment 0 map (right panel) for all H41$\alpha$-detected proplyds not detected in He41$\alpha$. All stacked He41$\alpha$ moment 0 maps are generated from velocity channels in the range $-120$ to $-90$ km s$^{-1}$ {with respect to the H41$\alpha$ rest frequency}, while all stacked H41$\alpha$ moment 0 maps are generated from velocity channels in the range $0$ to $60$ km s$^{-1}$ {with respect to the H41$\alpha$ rest frequency}. Contours show $3.5{\sigma}$, $4.5{\sigma}$, and $5.5{\sigma}$ emission. The synthesized beam is shown in the bottom left corner of each panel. { Stacked moment 0 maps are generated from our data cube with 10 km s$^{-1}$ channels.}
\label{fig:He41a_stack}}
\end{figure*}

Our stacking analysis indicates that the ONC contains a large population of proplyds with different physical conditions than the ones detected in this paper. 
Deeper radio recombination line observations are needed to 
determine whether it is the electron density, electron temperature, or gas metallicity that varies the most across photoevaporative disk winds in the ONC.

\subsection{Helium abundances}\label{sec:discussion:He}

{Previous studies have used hydrogen and helium radio recombination lines to measure y$^+$, the abundance of singly ionized helium relative to ionized hydrogen,  in {H\textsc{II}}  regions} 
\citep[e.g.,][]{Churchwell74, Shaver83, Peimbert88, Garay98, Roshi17, Pabst24}. Values of $y^+$ that exceed 0.08 are thought to arise from the enrichment of interstellar matter via galaxy evolution, since Big Bang nucleosynthesis suggests a primordial helium abundance of 0.08 for the Universe \citep{Churchwell74}. Values below 0.08 indicate that the helium gas is not fully ionized, 
which can occur when an {{H\textsc{II}}  region is excited by stars 
that do not emit intense levels of helium-ionizing ($\gtrsim24$ eV) radiation (i.e., spectral types later than ${\sim}$O6)}; when dust in the {H\textsc{II}}  region preferentially absorbs helium-ionizing photons; or, when non-LTE and/or opacity effects become significant \citep[see discussion in][]{Churchwell74, Garay98}. {When the helium and hydrogen gas are fully ionized, y$^+$ provides a direct measurement of the overall helium abundance 
\citep{Churchwell74}.}

The detection of H$41\alpha$ and  He$41\alpha$ towards 168-326 SE/NW enables us to {constrain} the helium abundances { of protoplanetary disks} with radio recombination lines.  
As shown in Table \ref{tab:source_properties_He}, we find that the He$41\alpha$ to H$41\alpha$ line ratio of 168-326 SE/NW implies a singly ionized helium abundance of ${\sim} 0.2$. This is somewhat larger than the typical helium abundances measured in {H\textsc{II}}  regions \citep{Churchwell74}. It is also larger than the expected helium abundance of the Orion Nebula, for which optical- and radio-recombination  line-based studies find 
values of ${\sim}0.1$ \citep[e.g.,][]{Baldwin91, Pabst24}.  

{
The enhanced helium abundance of 168-326 SE/NW may be driven by this system's 
proximity to $\theta^1$ Ori C. Typically, the helium abundance of the Orion Nebula is measured from observations targeting ionized gas that is ${\gtrsim} 0.05$ pc from $\theta^1$ Ori C \citep[e.g.,][]{Baldwin91, Pabst24}. 168-326 SE/NW is ${\sim}0.013$ pc (in projected separation) from $\theta^1$ Ori C (see Table \ref{tab:source_properties}), so it is likely exposed to stronger levels of helium-ionizing radiation and less affected by extinction 
than the gas at larger projected separations, in which case we might expect our derived helium abundance to be more reflective of the Orion Nebula's helium abundance. 
} 

{ 
Another possibility is that our helium abundance measurement is contaminated by line blending of helium and carbon recombination lines \citep[e.g.,][]{Churchwell74, Peimbert88, Pabst24}. The C41$\alpha$ recombination line is about ${\sim}8$ MHz (${\sim}30$ km s$^{-1}$) from the He41$\alpha$ line, and if C41$\alpha$ emission from 168-323 SE/NW were to have a similar line width as the detected H41$\alpha$ emission, it would partially overlap with He41$\alpha$.  The relative line fluxes of carbon and helium radio recombination lines vary depending on the incident radiation field, but in {H\textsc{II}}  regions excited ${\gtrsim}$O6 stars, carbon radio combination lines  are typically ${\sim} 5-10\times$ fainter than helium radio recombination lines \citep[for examples in the Orion Nebula, see][]{Pabst24}. Line blending may therefore only have a modest (${\sim}25\%$) effect on our helium abundance measurement for 168-326 SE/NW, but this needs to be confirmed with higher spectral resolution observations that isolate C41$\alpha$ and He41$\alpha$ emission. 
}

{An increased sample of proplyd helium abundance measurements, { combined with deeper and higher angular observations that can more readily spatially isolate He recombination lines from close-separation proplyds (i.e., 168-326 SE and 168-326 NW), can also help to}
determine whether proplyds are helium rich relative to the Orion Nebula or whether the Orion Nebula is more helium rich than currently thought.} Figure \ref{fig:He41a_stack} shows stacked average moment 0 maps of He$41\alpha$ and H$41\alpha$ for all H$41\alpha$-detected proplyds that are not detected in He$41\alpha$. We detect He$41\alpha$ at ${>}4\sigma$ in the stacked moment 0 maps, indicating that deeper observations will detect helium radio recombination lines from additional proplyds. We compute an average singly ionized helium abundance for the stacked sources by taking the ratio of the stacked He$41\alpha$ and H$41\alpha$ moment 0 emission. Depending on the chosen aperture, we find values of $y^+$ that range 
from ${\sim}0.12 - 0.16$. This suggests that the average H$41\alpha$-detected proplyd is less abundant in ionized helium than 168-326 SE/NW.  
The helium ionization fraction of proplyds\textemdash and by extension, gas in the Orion Nebula\textemdash may therefore vary depending on projected separation from $\theta^1$ Ori C.

{Deeper observations may also detect radio recombination lines from oxygen and other heavy atoms. These lines have recently been detected in the Orion Nebula  \citep[e.g.,][]{Liu23}, and if hydrogen, helium, carbon, and oxygen radio recombination lines were to be detected towards proplyds,} we would be able to measure the elemental abundances of photoevaporating disks with radio recombination lines. 
Since the most massive star in the ONC, $\theta^1$ Ori C, has a spectral type of O6 \citep{Odell17}, it is hot enough to emit the intense levels of hard (${>}10$ eV) EUV photons needed to ionize hydrogen, helium, and heavier atoms.  
This in contrast with other proplyd-hosting clusters in Orion \citep[e.g., NGC 1977 and NGC 2024;][]{Bally12, Kim16, Haworth21b, Boyden24}, whose most massive stars have later spectral types and are less likely to be emitting significant levels of hard EUV radiation \citep[e.g.,][]{Bik03, Peterson08}.

\section{\bf {Conclusions}}\label{sec:conclusions}

We presented ALMA observations that produced the first detection of radio recombination lines from a protoplanetary disk. We mosaicked the central $2\rlap{.}'0 \times 2\rlap{.}'5$ of the ONC at 3.1 mm, and because the ALMA spectral setup covered the $n = 42$ to $n = 41$ transitions of hydrogen and helium, we were able to search for hydrogen and helium radio recombination lines towards the positions of ${>}200$ protoplanetary disks. 

We detect H41$\alpha$ emission 
from 17 protoplanetary 
disks. 
The detections are all HST-identified proplyds within $0.06$ pc (in projection separation) of the massive star $\theta^1$ Ori C.  
For all detections, we find that the  H41$\alpha$ emission is associated with the locations of proplyd ionization fronts, indicating that the H41$\alpha$ line is tracing gas that has been externally photoevaporated off the disks' surfaces.  For a subset of detections, the H41$\alpha$ emission exhibits a near-linear velocity gradient that is consistent with the gradients expected from photoevaporative disk winds. 

We measured the line fluxes and line widths of the detected H41$\alpha$ lines by fitting Gaussian line profiles to the observed spectra. The H41$\alpha$-detected proplyds span a range of H41$\alpha$ line fluxes, but they typically have similar H41$\alpha$ line-to-continuum ratios. For most proplyds, the derived line-to-continuum ratios imply electron gas temperatures of ${\sim}7,000$ to ${\sim}9,000$ K {and electron densities of ${\sim}10^6$ to ${\sim}10^7$ cm$^{-3}$. We find that proplyd electron density 
correlates inversely with projected separation from $\theta^1$ Ori C, consistent with the behavior expected from a reduction in UV flux at increased projected separations. The derived electron temperatures, on the other hand, show no correlation with projected separation.}

For the H41$\alpha$ line widths, we find typical {full-width-at-half-maxima} of ${\sim}40-60$ km s$^{-1}$, which are significantly broader than the line widths expected from {thermal, turbulent, and pressure broadening.} We suggest that the broadening of proplyd H41$\alpha$ emission is dominated by { outflowing}  
gas motions associated with external photoevaporation, since line widths of ${\sim}40-60$ km s$^{-1}$ match up well with the expected terminal velocities of photoevaporative disk winds that have passed through an ionization front. However, for the couple of proplyds with line widths ${>}60$ km s$^{-1}$, we suspect that the presence of jets is also contributing to the overall line broadening.

Finally, we detect He41$\alpha$ emission 
from the H41$\alpha$-detected source 168-326 SE/NW. We find that this system's He41$\alpha$ to H41$\alpha$ line ratio implies a helium abundance that is greater than the {canonical}  helium abundance of the Orion Nebula. 
{However, with our current observations, we cannot rule out the possibility that our helium abundance measurement is contaminated by line blending of carbon and helium recombination lines.} 
{Deeper, higher spectral resolution observations are needed to more reliably measure the helium abundances of proplyds.} 
{Our stacking analysis suggests that the helium ionization fraction of proplyds correlates with projected separation from $\theta^1$ Ori C, } but this needs to be confirmed with direct measurements.

Our study demonstrates that radio recombination  lines are readily detectable in { ionized externally photoevaporating disks}. 
This essentially opens up a new way 
to  
measure disk properties in clustered (i.e., typical) star-formation environments. 
{Millimeter- and submillimeter-band} 
recombination lines appear especially suitable for mapping the kinematics of 
photoevaporating disks, given that these lines {are less likely to be affected}
by pressure broadening. Millimeter- and centimeter-band 
recombination lines appear suitable for measuring the densities, temperatures, and chemistry of 
{proplyd ionization fronts,}
since these lines remain in LTE down to low gas densities and temperatures.  {In the immediate future, we intend to develop a framework for measuring the mass loss rates of ionized photoevaporating disks with high angular resolution radio recombination line observations.}

{ By measuring the physical conditions of large samples of} 
photoevaporating disks with radio recombination lines, we will be able to gain new insights into how the radiation environments of stellar clusters shape the demographics of disks and exoplanets. 
{When ALMA’s Wideband Sensitivity Upgrade is completed, and when the next-generation Very Large Array is commissioned, we expect radio recombination line surveys of protoplanetary disks to become increasingly feasible over large disk samples.}

\vspace{0.1in}

{\it \noindent  Acknowledgements:} 
We are grateful to NRAO staff, especially E. Starr, for providing assistance with imaging the ALMA line observations. We also acknowledge the use of NRAO computing facilities for the reduction and imaging of the ALMA data. 
R. Boyden acknowledges support from the the Virginia Initiative on Cosmic Origins (VICO) and NSF grant no. AST-2206437. 
{N. Ballering acknowledges support from NSF grant no. AST-2205698, and NASA/Space Telescope Science Institute grant JWST-GO-03271.
Support for C.J.L. was provided by NASA through the NASA Hubble Fellowship grant No. HST-HF2-51535.001-A awarded by the Space Telescope Science Institute, which is operated by the Association of Universities for Research in Astronomy, Inc., for NASA, under contract NAS5-26555.
T. Haworth acknowledges UKRI guaranteed funding for a Horizon Europe ERC consolidator grant (EP/Y024710/1) and a Royal Society Dorothy Hodgkin
Fellowship. 
{ J.C.T. acknowledges support from NSF grants AST-2009674 and AST-2206450 and ERC
Advanced grant MSTAR.
L.I.C. acknowledges support from the David and Lucille Packard Foundation, Research Corporation for Science Advancement Cottrell Fellowship, NASA ATP 80NSSC20K0529, and NSF grant no.\ AST-2205698. 
Z.-Y. Li is supported in part by NASA 80NSSC20K0533 and NSF AST-2307199.
J.C.T., L.I.C., and Z-Y. L. acknowledges funding from the Virginia Institute for Theoretical Astrophysics (VITA), supported by the College and Graduate School of Arts and Sciences at the University of Virginia.}
}

This paper makes use of the following ALMA data: ADS/JAO.ALMA\#2015.1.00534.S and ADS/JAO.ALMA\#2018.1.01107.S.  
ALMA is a partnership of ESO (representing its member states), NSF (USA) and NINS (Japan), together with NRC (Canada), MOST and ASIAA (Taiwan), and KASI (Republic of Korea), in cooperation with the Republic of Chile. The Joint ALMA Observatory is operated by ESO, AUI/NRAO and NAOJ. 
The National Radio Astronomy Observatory and Green Band Observatory are facilities of the U.S. National Science Foundation operated under cooperative agreement by Associated Universities, Inc.
Some of the data presented in this paper were obtained from the Mikulski Archive for Space Telescopes (MAST) at the Space Telescope Science Institute. The specific observations analyzed can be accessed via doi:10.17909/1cjq-yt93.

{\it Facilities: Atacama Large Millimeter/submillimeter Array (ALMA), NSF's Karl G. Jansky Very Large Array (VLA)}

{\it Software:} {\tt Astropy} \citep{astropy13, Astropy18, Astropy22}, 
{\tt CASA} \citep{CASA22}, {\tt matplotlib} \citep{Hunter07},
{\tt CARTA} \citep{Comrie21}, {\tt pyspeckit} \citep{Ginsburg22}

\vspace{-0.15in}

\bibliography{Boyden_bib}




\appendix



\section{Continuum Images}\label{appendix:dust}

\begin{figure*}[ht!]
    \epsscale{1.2}
    \hspace{-0.3in}    \plotone{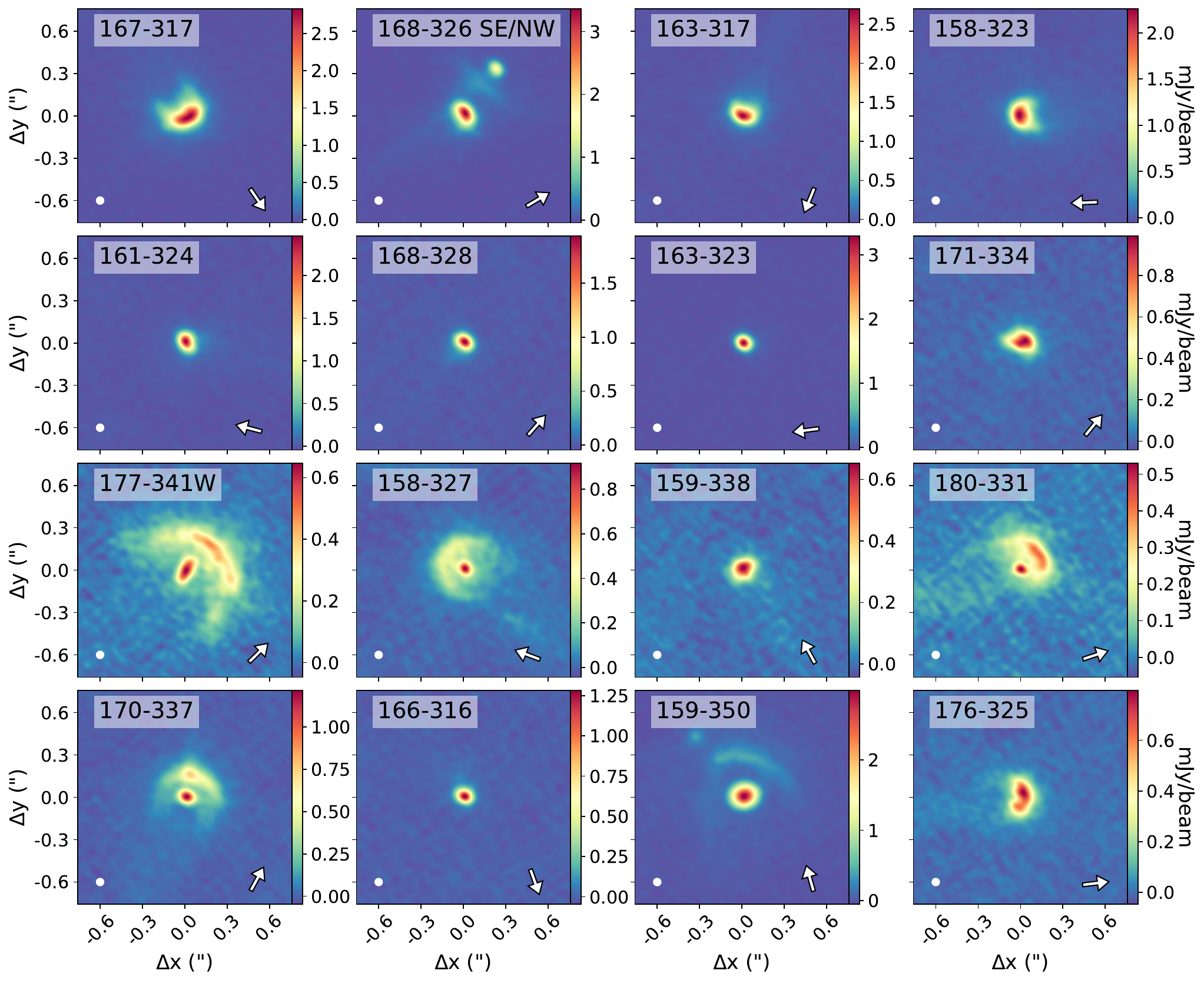}
     \vspace{-0.1in}
\caption{3.1 mm continuum images of H41$\alpha$-detected proplyds in the ONC. Each image is generated as a sub-image from the larger 3.1 mm continuum mosaic from \cite{Ballering23}. The synthesized beam is shown in the botton left corner of each panel.  The arrow in each panel points to the direction of $\theta^1$ Ori C. 
\label{fig:dust}}
\end{figure*}

{  
Here we include Figure \ref{fig:dust}, which shows the 3.1 mm continuum sub-images of each H41$\alpha$-detected proplyd. These sub-images are generated from the full high-resolution continuum mosaic of \cite{Ballering23}. 
} 

\clearpage

\twocolumngrid

\section{Additional Spectra Plots}\label{appendix:Voigt}

\begin{figure}[ht!]
    \epsscale{1.2}
    \hspace{-0.3in}    \plotone{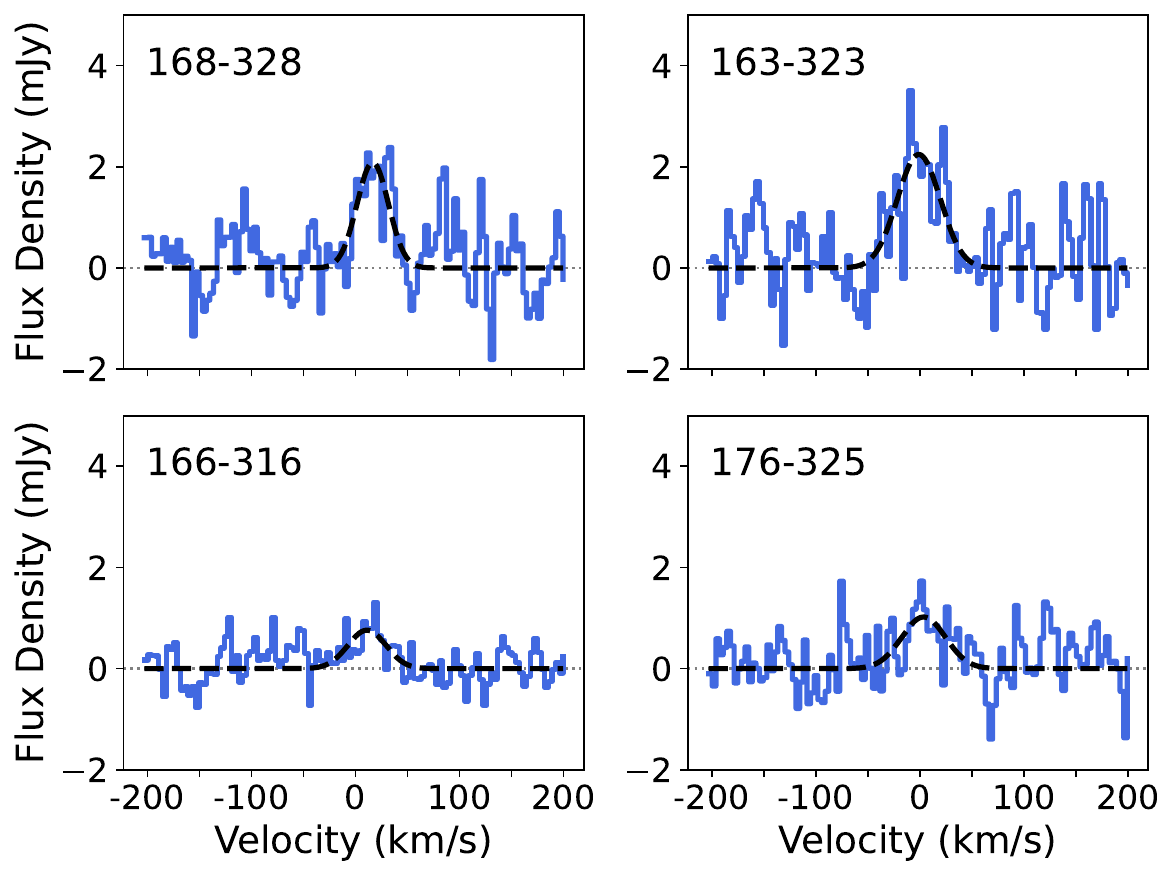}
      \vspace{-0.1in}
\caption{Zoomed-in radio recombination line spectra (blue lines) and best-fit Gaussian line profiles (black lines) for proplyds 168-328, 163-323, 166-316, and 176-325. These spectra are also shown in Figure \ref{fig:spec_detections}, expect here we use smaller y-axis limits. 
\label{fig:spec_detections_zoom}}
\end{figure}


{ Here we include Figure \ref{fig:spec_detections_zoom}, which shows zoomed-in radio recombination line spectra for proplyds 168-328, 163-323, 166-316, and 176-325.} 

\begin{figure}[ht!]
    \epsscale{1.0}
    \hspace{-0.3in}    \plotone{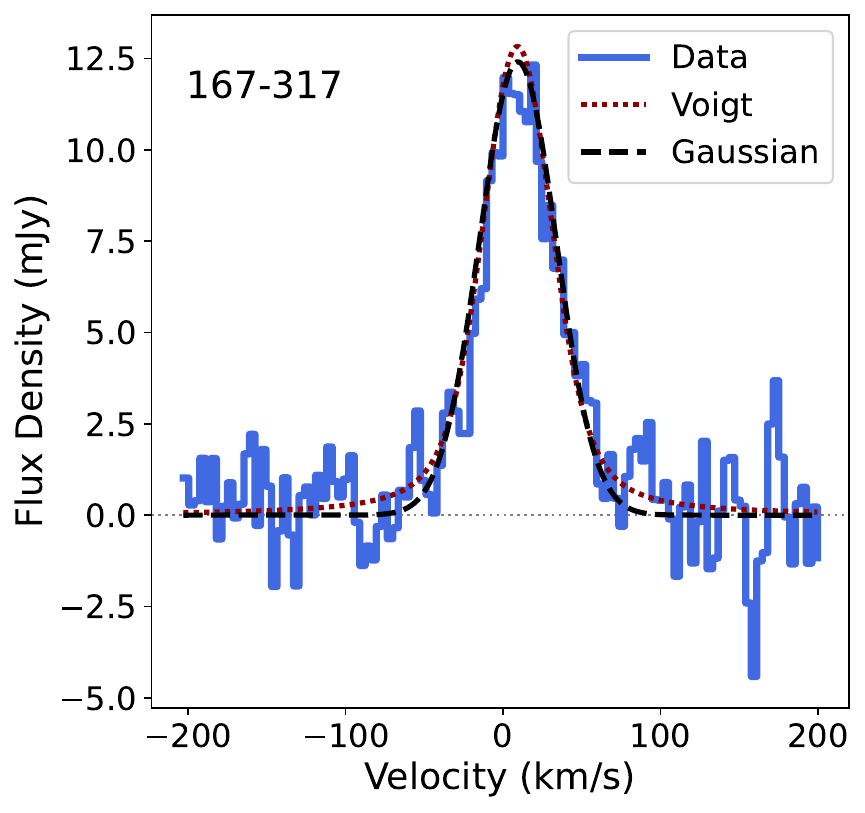}
      \vspace{-0.1in}
\caption{Radio recombination line spectra (solid blue line), best-fit Gaussian line profile (dashed black line), and best-fit Voigt line profile (dotted brown line) for proplyd 167-317. 
\label{fig:voigt}}
\end{figure}

{  
We also include Figure \ref{fig:voigt}, which plots best-fit Gaussian and Voigt profiles for the highest signal-to-noise detection in our sample, proplyd 167-317. Gaussian and Voigt profile fitting are both performed using {\tt pyspeckit} \citep{Ginsburg22}.
Both fits yield similar reduced $\chi^2$ values, but differ in the velocity channels that are away from the best-fit systemic velocities and where the signal-to-noise ratio is the lowest.
}

\end{document}